\documentclass{article} 
\usepackage{epsfig} 
\usepackage{cite}  
\def\CA{C_A}

\def\NF{N_F}
\def\Poles{{\cal P}oles}
\def\Finite{{\cal F}inite}
\def\bom#1{{\mbox{\boldmath $#1$}}}
\def\MSbar{$\overline{{\rm MS}}$}
\def\ket#1{|{#1}\rangle}
\def\bra#1{\langle{#1}|}
\def\braket#1#2{\langle #1 |#2 \rangle}
\def\cm{{\cal M}}

\def\e{\epsilon}
\def\d{\hbox{d}}

\begin{document} 
\unitlength1cm 
\begin{titlepage} 
\vspace*{-1cm} 
\begin{flushright} 
ZU--TH 03/04\\
IPPP/04/12\\
DCPT/04/24\\
NSF-KITP-04-32\\
hep-ph/0403057\\
March 2004
\end{flushright} 
\vskip 2.5cm 

\begin{center} 
{\Large\bf Infrared Structure of $e^+e^- \to 2$~jets at NNLO}
\vskip 1.cm 
{\large  A.~Gehrmann--De Ridder}$^{a,c}$, {\large  T.~Gehrmann}$^{a,c}$ 
and {\large E.W.N.~Glover}$^{b,c}$ 
\vskip .7cm 
{\it $^a$ Institut f\"ur Theoretische Physik, Universit\"at Z\"urich,
Winterthurerstrasse 190,\\ CH-8057 Z\"urich, Switzerland} 
\vskip .4cm 
{\it $^b$ Institute for Particle Physics Phenomenology, University of Durham,
South Road,\\ Durham DH1 3LE, England} 
\vskip .4cm 
{\it $^c$ Kavli Institute for Theoretical Physics, University of California,\\
Santa Barbara, CA 93106, USA} 
\end{center} 
\vskip 2.6cm 

\begin{abstract} 
The production of two jets is the simplest exclusive quantum chromodynamics 
process in electron-positron annihilation. Using this process, we examine the
structure of next-to-next-to-leading order (NNLO) corrections to jet production
observables. We derive a subtraction formalism including double real radiation
at tree level and single real radiation at one loop. For two-jet production, 
these subtraction terms coincide with the full matrix elements, thus
highlighting the phase space structure of the subtraction procedure.  We then
analytically compute the infrared singularities arising from  each partonic
channel. For the purely virtual (two-parton) NNLO corrections, these take the
well known form predicted by Catani's infrared factorization formula.  We
demonstrate that individual terms in the infrared factorization formula can be
identified with infrared singular terms from three- and four-parton  final
states, leaving only single poles and a contribution from  the one-loop soft
gluon current, which subsequently cancels between the three- and four-parton 
final states. Summing over all different final states, we observe an explicit 
cancellation of all infrared poles and recover the known two-loop correction 
to the hadronic $R$-ratio. 
\end{abstract} 
\vfill 
\end{titlepage} 
\newpage 

\renewcommand{\theequation}{\mbox{\arabic{section}.\arabic{equation}}} 

\section{Introduction}
\setcounter{equation}{0}

Jet production observables can be measured very accurately at present  high
energy colliders. By confronting these data with theoretical calculations, one
can determine the strong coupling constant or parton distribution
functions~\cite{dissertori}.  Analyzing the different sources of error on these
determinations, it becomes clear that the largest source of uncertainty are
currently the insufficiently precise theoretical calculations of QCD
corrections to jet observables, which mostly include terms only to the
next-to-leading order (NLO) in perturbation theory. 

To improve this situation, the calculation of next-to-next-to-leading order
(NNLO) corrections to jet observables  is mandatory. For an $n$-jet observable,
several ingredients are required; the two-loop $n$-parton matrix elements,  the
one-loop ($n$+1)-parton matrix elements and the tree level ($n$+2)-parton 
matrix elements. In the recent past, enormous progress has been made  in the
calculation of two-loop $2\to 2$ and $1\to 3$  QCD matrix elements, which are
now known for all  massless parton--parton scattering 
processes~\cite{m1,m2,m3,m4} relevant to hadron colliders as well as for 
$\gamma^*\to q\bar q g$~\cite{3jme,muw2} and its crossings~\cite{ancont}.  For
the corresponding partonic processes,  the one-loop matrix elements with one 
additional parton~\cite{5p,V4p}  and the tree-level matrix elements with two
more partons  are also known and form part of NLO programs for  $1\to
4$~\cite{nplusone1,cullen} and $2\to 3$ reactions~\cite{nplusone2}.  Since these
matrix elements lead to infrared singularities due to one or two partons
becoming theoretically unresolved (soft or collinear), one needs to find one-
and two-particle subtraction terms which account for these singularities  in
the matrix elements, and which are sufficiently simple to be integrated
analytically over the unresolved phase space. One-particle subtraction at tree
level is well understood from NLO  calculations\cite{ert,kramer,eks,gg,cs}  and
general algorithms are available for  one-particle subtraction at one
loop\cite{onel,cg,kos1,weinzierl2}, in a  form that has recently been
integrated  analytically~\cite{kos1,weinzierl2}.

Tree-level two-particle subtraction terms have been extensively studied in the 
literature\cite{twot,kosower,weinzierl1}. However  their integration over the
unresolved phase space remains an outstanding  issue. The problem of
integrating out double real emission contributions  has so far only been
addressed in specific  calculations~\cite{ggam,uwer,fg,daleo}, each of which 
requires a subset of the ingredients needed for generic jet observables at
NNLO.

In this paper, we aim to contribute to a formulation of  a general subtraction
scheme including double real emission by studying the  simplest QCD process,
two-jet production in $e^+e^-$ annihilation at NNLO.  
Using the iterated sector decomposition~\cite{hepp,itsec}
for the treatment  of all real emission 
contributions~\cite{itsec1,ggh,babis} and
without the need to define  subtraction terms, the NNLO
corrections to this process have been obtained in a numerical
form~\cite{babisnew} very recently. 

In this work,  we extract the infrared
divergent parts of all partonic contributions to two-jet final states
analytically and identify them with known structures in the virtual two-loop
~\cite{catani} and one-loop single unresolved~\cite{cg} corrections. These
identifications might help to construct a general NNLO subtraction formalism,
which would allow the NNLO computation of three-jet production in 
$e^+e^-$-annihilation and, in the longer term,  the inclusion of  subtraction
terms for double initial state emission, as required for the NNLO corrections
to jet observables at lepton-hadron and hadron-hadron colliders. 

This paper is organized as follows. In Section~\ref{sec:contributions},  we
describe the different partonic contributions yielding two-jet final states  at
NNLO. Section~\ref{sec:irsub} describes a formalism to subtract  the infrared
singularities from single and double real emission present in these
contributions. All partonic channels are computed in Section~\ref{sec:nnlo} 
and decomposed into infrared finite and infrared divergent parts, the latter
being identified with known structures. The integrated partonic terms 
represent the analytic integrals of the subtraction terms  accounting for all
real singularities. The structure of the infrared cancellations is illustrated
in Section~\ref{sec:ircancel}, where we also check the correctness of our
results by rederiving the NNLO corrections to the  hadronic $R$-ratio. An
appendix listing all integrals needed in this calculation is enclosed.

\section{Perturbative corrections to two-jet final states}
\label{sec:contributions}
\setcounter{equation}{0}
Two-jet final states in $e^+e^-$ annihilation are produced 
by the primary process 
$\gamma^{*} \to q \bar{q}$, the decay of a virtual photon into a
 quark--antiquark
pair,
\begin{equation}
\gamma^* (q) \longrightarrow q(p_1) + \bar q (p_2)\; .
\end{equation}

At higher orders in perturbation theory, this process receives 
corrections from the exchange of virtual particles and 
from real radiation. While the NLO corrections involve up to 
three final state partons, one finds at NNLO that 
two-jet final states are produced through partonic final states 
including up to four partons. The individual 
partonic channels are:\\

\begin{tabular}{lll}
LO & $\gamma^*(q)\to q(p_1)\,\bar q(p_2)$ & tree level \\[2mm]
NLO & $\gamma^*(q)\to q(p_1)\,\bar q(p_2)$ & one loop \\
 & $\gamma^*(q)\to q(p_1)\,\bar q(p_2)\, g(p_3)$ & tree level \\[2mm]
NNLO & $\gamma^*(q)\to q(p_1)\,\bar q(p_2)$ & two loop \\
 & $\gamma^*(q)\to q(p_1)\,\bar q(p_2)\, g(p_3)$ & one loop \\
& $\gamma^*(q)\to q(p_1)\,\bar q(p_2)\, q'(p_3)\,\bar q'(p_4)$ & tree level \\
& $\gamma^*(q)\to q(p_1)\,\bar q(p_2)\, q(p_3)\,\bar q(p_4)$ & tree level \\
& $\gamma^*(q)\to q(p_1)\,\bar q(p_2)\, g(p_3)\,g(p_4)$ & tree level\\ 
\end{tabular}
\vspace{2mm}

The contributions of these processes are weighted by 
so-called jet functions (described in detail in Section~\ref{sec:irsub}
below), which select two-jet final states from the partonic final
state momenta. For the analytic extraction of infrared poles present in 
all partonic channels, we will determine the inclusive $i$-parton production
cross sections for these processes. They are 
described by a single Lorentz-invariant
\begin{equation}
q^2 = \left(\sum_i p_i\right)^2\;.
\end{equation}

The $n$-particle phase space in 
dimensional regularization with $d=4-2\e$ space-time dimensions reads,
\begin{equation}
\d \Phi_n = \frac{\d^{d-1} p_1}{2E_1 (2\pi)^{d-1}}\; \ldots \;
\frac{\d^{d-1} p_n}{2E_n (2\pi)^{d-1}}\; (2\pi)^{d} \;
\delta^d (q - p_1 - \ldots - p_n) \,.
\end{equation}
Parameterizations of the three- and four-particle massless phase space 
appropriate to analytic integration are discussed in~\cite{ggh}. In the
following, we frequently
normalize the three- and four-particle phase space to the volume of the 
two-particle phase space,
\begin{equation}
P_2  = \int \d \Phi_2 =
2^{-3+2\e}\, \pi^{-1+\e}\, \frac{\Gamma(1-\e)}{\Gamma(2-2\e)}\,
 (q^2)^{-\e} \; .
\end{equation}

\section{Infrared subtraction terms}
\setcounter{equation}{0}
\label{sec:irsub}
To obtain the perturbative corrections to a jet observable at a given  order,
all partonic multiplicity channels contributing to this order  have to be
summed. Each partonic channel contains infrared singularities. However, after
summation, all singularities cancel among each other~\cite{kln}.

While infrared singularities from purely virtual corrections are obtained 
immediately after integration over the loop momenta, their extraction is  more
involved for real emission (or mixed real-virtual) contributions. Here,
the infrared singularities become only explicit after integrating  the real
radiation matrix elements over the phase space appropriate to  the jet
observable under consideration. In general, this integration  involves the
(often iterative) definition of the jet observable, such that  an analytic
integration is not feasible (and also not appropriate). Instead,   one would
like to have a flexible method that can be easily adapted to  different jet
observables. Therefore, the infrared singularities  of the real radiation
contributions should be extracted using  infrared subtraction  terms. These
subtraction terms are constructed such that they approximate the full  real
radiation matrix elements in all singular limits while still being 
sufficiently simple to be integrated analytically over a section of  phase
space that encompasses all regions corresponding to singular configurations.

To specify the notation, we define 
the tree level $m$-parton contribution to the $J$-jet cross section 
in $d$-dimensions by,
\begin{equation}
{\rm d} \sigma^{B}=N_{{in}}\sum_{{m}}{\rm d}\Phi_{m}(p_{1},...,p_{m},Q)
\frac{1}{S_{{m}}}\,|{\cal M}_{m}(p_{1},...p_{m})|^{2}\; 
{\cal F}_{J}^{(m)}(p_{1},...,p_{m}).
\label{eq:sigm}
\end{equation}
${\cal N}_{in}$ includes all QCD-independent factors,
$\sum_{m}$ denotes the sum over all configurations with $m$ partons,
${\rm d}\Phi_{m}$ is the phase space for $m$ partons, $S_{m}$ is a
symmetry factor for identical partons in the final state 
and finally $|{\cal M}_{m}|$ is the tree level $m$-parton matrix element.
The jet function $ {\cal F}_{J}^{(m)}$ defines the procedure for 
building $J$-jets out of $m$ partons.
The main property of ${\cal F}_{J}^{(m)}$ is that the jet observable defined
above is collinear and infrared safe as explained in \cite{cs}.
In general ${\cal F}_{J}^{(m)}$ contains $\theta$ and $\delta$-functions.
and ${\cal F}_{2}^{(2)}=1$. 

\subsection{NLO infrared subtraction terms}

At NLO, we consider the following $m$-jet cross section,
\begin{equation}
{\rm d}\sigma_{NLO}=\int_{{\rm d}\Phi_{m+1}}\left({\rm d}\sigma^{R}_{NLO}
-{\rm d}\sigma^{S}_{NLO}\right) +\left [\int_{{\rm d}\Phi_{m+1}}
{\rm d}\sigma^{S}_{NLO}+\int_{{\rm d}\Phi_{m}}{\rm d}\sigma^{V}_{NLO}\right].
\end{equation}
The cross section ${\rm d}\sigma^{R}_{NLO}$ has the same expression as the 
Born cross section ${\rm d}\sigma^{B}_{NLO}$ (\ref{eq:sigm}) above
except that $m \to m+1$, while 
${\rm d}\sigma^{V}_{NLO}$ is the one-loop virtual correction to the 
$m$-parton Born cross section ${\rm d}\sigma^{B}$.
The cross section ${\rm d}\sigma^{S}_{NLO}$ is a local counter-term for  
 ${\rm d}\sigma^{R}_{NLO}$. It has the same unintegrated
singular behavior as ${\rm d}\sigma^{R}_{NLO}$ in all appropriate limits.
Their difference is free of divergences 
and can be integrated over the $(m+1)$-parton phase space numerically.
The subtraction term  ${\rm d}\sigma^{S}_{NLO}$ has 
to be integrated analytically over all singular regions of the 
$m+1$-parton phase space. 
The resulting cross section added to the virtual contribution 
yields an infrared finite result. 

The subtraction term  ${\rm d}\sigma^{S}_{NLO}$
can be constructed in a number of different ways, and even the phase space 
used with these subtraction terms can vary from method to method. In 
general, two different methods are applied at next-to-leading order. In the 
{\it slicing method}~\cite{kramer,gg}, 
one exploits the fact that both real radiation matrix element 
and final state phase space factorize in all soft and collinear limits. 
Therefore, a subtraction term for a given singular limit 
can be obtained by simply expanding the full 
matrix element around the limit under consideration. This subtraction term 
is then integrated over a small slice of phase space, which is tailored to 
contain only one singular configuration. In the {\it subtraction 
method}~\cite{ert,eks,cs}, the infrared subtraction terms are 
integrated over the whole final state phase space (or at least 
over the full 
sub-phase space appropriate to one or more final state momenta).  In this 
method, care has to be taken in order to construct subtraction terms 
which fully account for the limit they are aimed at without introducing 
spurious infrared singularities in other limits. 

A systematic procedure for finding NLO infrared subtraction terms in the 
second method is the dipole formalism 
derived by Catani and Seymour~\cite{cs}.
Their subtraction terms are obtained as 
 sum of dipoles $\sum {\cal D}_{ijk}$ (where each dipole corresponds to a
single infrared singular configuration)
such that,
\begin{eqnarray}
{\rm d}\sigma_{NLO}^{R}-{\rm d}\sigma_{NLO}^{S}
&= & N_{in}\sum_{m+1}{\rm d}\Phi_{m+1}(p_{1},...,p_{m+1},Q)
\frac{1}{S_{{m+1}}} \,\Bigg [|{\cal M}_{m+1}(p_{1},...p_{m+1})|^{2}\;
{\cal F}_{J}^{(m+1)}(p_{1},...,p_{m+1}) \nonumber \\
&&-\sum_{{\rm pairs~} i,j}\;\sum_{k \neq i,j}{\cal D}_{ijk}\,
|{\cal M}_{m}((p_{1},..\tilde{p}_{ij},\tilde{p}_{k},...,p_{m+1})|^2\,
{\cal F}_{J}^{(m)}(p_{1},..\tilde{p}_{ij},\tilde{p}_{k},...,p_{m+1})\;\Bigg ].
\label{eq:sub1}
\end{eqnarray}
The dipole contribution ${\cal D}_{ijk}$ involves the $m$-parton amplitude 
depending only on the redefined on-shell momenta
$p_{1},..\tilde{p}_{ij},\tilde{p}_{k},p_{m+1}$ 
and the splitting matrix $V_{ij,k}/s_{ij}$ which depends only on 
$p_{i},p_{j},{p}_{k}$.
The momenta ${p}_{i}$, $p_j$ and ${p}_{k}$ are respectively 
the emitter, unresolved parton  and the spectator momenta 
corresponding to a single dipole term. The redefined on-shell momenta 
$  \tilde{p}_{ij},\tilde{p}_{k}$ are linear combinations of them.

For the NLO two-jet cross section, the real corrections involve 
three partons in the final state. 
The single unresolved configurations
occur when the invariants $s_{13}$ or $s_{23}$ 
vanish. We notice that the subtraction term built
as the sum of the dipoles is very close to 
the full three-parton matrix element. 
This argument can in fact be turned around: the dipole subtraction terms 
can be obtained by an appropriate partial fractioning of the full matrix
element squared into terms which can be uniquely attributed 
to individual single unresolved configurations. In the following, we shall 
use this formulation as {\it definition} of a dipole term, without 
specifying its precise form. 

The jet function ${\cal F}^{(m)}_J$ in (\ref{eq:sub1}) depends not on the 
individual momenta ${p}_{i}$, $p_j$ and ${p}_{k}$, but only on 
$\tilde{p}_{ij},\tilde{p}_{k}$. One can therefore carry out the integration
over the unresolved dipole phase space appropriate to 
   ${p}_{i}$, $p_j$ and ${p}_{k}$ analytically, exploiting the 
dipole factorization of the phase space~\cite{cs},
\begin{equation}
\d \Phi_{m+1}(p_{1},...,p_{m+1},Q)  = 
\d \Phi_{m}(p_{1},..\tilde{p}_{ij},\tilde{p}_{k},...,p_{m+1})\cdot 
\d \Phi_D (p_i,p_j,p_k)\;,
\end{equation}
where $\d \Phi_D $ is symmetric under permutations of $i,j,k$.
The dipole phase space is proportional to the three-particle phase space,
as can be seen by using $m=2$ in the above formula and exploiting 
the fact that the two-particle phase space is a constant,
\begin{equation}
P_2 = \int \d \Phi_2\;,
\end{equation}
such that
\begin{equation}
{\rm d}\Phi_{3}= P_2\; {\rm d}\Phi_{D}\;.
\end{equation}

It should be noted in this context, that  the sum of dipoles is much easier to
integrate over the dipole phase space  than an individual dipole.   In what
follows, we shall  take into account this fact, and choose to use  the full
three-parton matrix element as subtraction term~\cite{cullen}
 to the real corrections to the
two-jet rate at NLO (and, in Section~\ref{sec:nnlosub} below, the full 
four-parton matrix element as candidate subtraction term  at NNLO). Taking the full
matrix element as the subtraction term provides a clear
visualization of the phase space regions covered by the different 
contributions involving the jet-functions ${\cal F}_{J}^{(m)}$. At NLO the
difference of the unintegrated real and subtracted  contributions to the two-jet
cross section reads,
\begin{eqnarray}
{\rm d}\sigma_{NLO}^{R}-{\rm d}\sigma_{NLO}^{S} &=&
N_{in}\;{\rm d}\Phi_{3}(p_{1},...,p_{3},Q)\;
|{\cal M}_{3}(p_{1},...p_{3})|^{2}\;\times \nonumber \\
& &\left( 
{\cal F}_{2}^{(3)}(p_{1},p_{2},p_{3})\;
-\frac{1}{2}{\cal F}_{2}^{(2)}(\widetilde{p}_{13},\tilde{p}_{2}) 
-\frac{1}{2}{\cal F}_{2}^{(2)}(\widetilde{p}_{23},\tilde{p}_{1}) \right).
\label{eq:NLOirsub}
\end{eqnarray}
The factors $1/2$ in front of the ${\cal F}_{2}^{(2)}$ are 
obtained by first writing out both dipoles, 
i.e.\ both terms obtained by partial fractioning of the three-parton matrix element which are appropriate to the 
configurations where parton 3 is unresolved with respect to either
parton 1 or 2.  Exploiting the fact that ${\cal F}_{2}^{(2)}=1$, 
these dipoles can be added together to yield the full 
matrix element. Reintroducing both jet functions consequently yields a 
factor $1/2$ for each of the momentum redefinitions. 

Equation~(\ref{eq:NLOirsub}) can be interpreted as follows:
out of the three-parton final state phase space one considers 
the two-jet configuration (defined by ${\cal F}_{2}^{(3)}$) 
minus the fully inclusive phase space (defined by ${\cal F}_{2}^{(2)}=1$), 
which covers to this order the two- and three-jet configurations. 

For the analytic integration, 
the subtraction term is rewritten as, 
\begin{equation}
{\rm d}\sigma^{S}_{NLO}={\rm d}\Phi_{3}\;|{\cal M}_{3}|^2\ = 
{\rm d}\Phi_{2}\;|{\cal M}_{2}|^2\;
\int_{{\rm d} \Phi_{D}}\;|M_{3}|^2\;,
\end{equation}
where the three-parton phase space is now a product 
of the dipole phase space and a two-parton phase space and where the 
three-parton matrix element is normalized to the two-parton matrix element 
such that
\begin{equation}
|M_{j}|^2 \equiv \frac{1}{|{\cal M}_{2}|^2} \, |{\cal M}_{j}|^2\;.
\end{equation}
The analytic integral of the subtraction term is 
therefore the three-parton contribution integrated over the fully inclusive 
phase space. The three-parton contribution to two-jet final states is
thus obtained by adding and subtracting the three-parton contribution 
to the inclusive cross section. The subtracted term is then used in the
numerical integration, accounting for all infrared 
singularities in the two-jet final states, while the added term is integrated 
analytically to make the infrared singularities explicit.

To summarize schematically the contributions 
to the two-jet cross section at NLO, one finds
the following two- and three-parton phase space contributions
\begin{equation}
{\rm d}\sigma^{2j}_{{\rm NLO}}=
{\rm d}\Phi_{3}\;|{\cal M}_{3}(p_{1},...p_{3})|^{2}\,
\left[({\cal F}_{2}^{(3)}-{\cal F}_{2}^{(2)})\right]\\
+{\rm d}\Phi_{2}\;|{\cal M}_{2}|^2 \;{\cal F}_{2}^{(2)}
\left(\int_{{\rm d}\Phi_{D}}|M_{3}|^2 + |M_{2}^{V,1}|^2 \right),
\label{eq:nlosum}
\end{equation}
where 
$|M_{j}^{V,1}|^2$ is the normalized one-loop virtual correction to the 
 $j$-parton matrix element, which exactly cancels all infrared poles  
obtained from the integral of the subtraction term. 
Both virtual and subtraction terms are proportional to the two-parton 
matrix element and phase space. 

\subsection{NNLO infrared subtraction terms}
\label{sec:nnlosub}
At NNLO, the two-jet production is induced by final states containing up to
four partons, including the one-loop virtual corrections to three-parton final 
states. As at NLO, one has to introduce subtraction terms for the 
three- and four-parton contributions. 
Schematically the NNLO two-jet cross section reads,
\begin{eqnarray}
{\rm d}\sigma_{NNLO}&=&\int_{{\rm d}\Phi_{4}}\left({\rm d}\sigma^{R}_{NNLO}
-{\rm d}\sigma^{S}_{NNLO}\right) + \int_{{\rm d}\Phi_{4}}
{\rm d}\sigma^{S}_{NNLO}\nonumber \\ 
&&+\int_{{\rm d}\Phi_{3}}\left({\rm d}\sigma^{V,1}_{NNLO}
-{\rm d}\sigma^{VS,1}_{NNLO}\right)
+\int_{{\rm d}\Phi_{3}}{\rm d}\sigma^{VS,1}_{NNLO}  
+ \int_{{\rm d}\Phi_{2}}{\rm d}\sigma^{V,2}_{NNLO}\;,
\end{eqnarray}
where $\d \sigma^{S}_{NNLO}$ denotes the real radiation subtraction term 
coinciding with the four-parton tree level cross section 
 $\d \sigma^{R}_{NNLO}$ in all singular limits. 
Likewise, $\d \sigma^{VS,1}_{NNLO}$
is the one-loop virtual subtraction term 
coinciding with the one-loop three-parton cross section 
 $\d \sigma^{V,1}_{NNLO}$ in all singular limits. 
Finally, the two-loop correction 
to the two-parton cross section is denoted by ${\rm d}\sigma^{V,2}_{NNLO}$.

A method to construct subtraction terms for $n$-parton cross sections at NNLO
has been worked out in  detail by Weinzierl for double real
radiation~\cite{weinzierl1} and  single real radiation off one-loop matrix
elements~\cite{weinzierl2}.  The method which we will outline in the following
differs from  that proposed in~\cite{weinzierl1,weinzierl2}, and we shall
highlight the main differences at the end of this section. 

Following the line of thought elaborated 
in the NLO discussion above, the subtraction terms are chosen to be the full
matrix elements, i.e.\ the sum of the full four-parton matrix 
elements related to the tree level processes 
$\gamma^{*} \to q \bar q g g$, $\gamma^{*} \to q \bar q q \bar{q}$ 
and $\gamma^{*} \to q \bar q q' \bar{q'} (q\neq q')$ for $\d \sigma^{S}$
and the one-loop corrected matrix element squared 
of $\gamma^{*} \to q \bar{q} g$ for  $\d \sigma^{VS,1}_{NNLO}$.

Clearly, these subtraction terms have the appropriate behavior 
in all singular limits. Moreover, as explained 
in \cite{ggh}, they  can be integrated analytically by reducing them 
algebraically to four master integrals which have been derived explicitly 
in~\cite{ggh}.
Likewise 
the integration of the one-loop correction to 
$ \gamma^{*} \to q \bar{q} g$ over the dipole phase space  can be performed 
analytically by reducing it to three master integrals which are 
listed in the appendix. These calculations are presented in 
Section~\ref{sec:nnlo} below. 

To perform the phase space integration of the subtraction terms, one 
needs to factorize  the phase space into a 
product of a two-particle phase space multiplying a dipole phase space
for the three-particle contribution and multiplying a tripole phase
space 
\begin{equation}
{\rm d}\Phi_{4} =  P_{2}\,{\rm d}\Phi_{T},
\label{eq:triphase}
\end{equation}
for the four-particle contribution. 
The latter factorization~\cite{kosower,uwer,ggh} is obtained by 
redefining a set of four massless on-shell momenta (emitter, two unresolved
partons, spectator) into two on-shell massless momenta. 

The construction of the NNLO subtraction term  to the four-parton tree level 
cross section  starts from the full four-parton matrix element:
\begin{equation}
{\rm d}\sigma_{NNLO}^{S,0} =
N_{in}\;{\rm d}\Phi_{4}(p_{1},...,p_{4},Q)\;
|{\cal M}_{4}(p_{1},...p_{4})|^{2} \left(
\frac{1}{2}{\cal F}_{2}^{(2)}(\tilde{p}_{134},\tilde{p}_{2}) 
+\frac{1}{2}{\cal F}_{2}^{(2)}(\tilde{p}_{234},\tilde{p}_{1}) 
\right)\; ,
\end{equation}
which is in analogy to the NLO subtraction term.
However, in order to have a subtraction term 
which correctly accounts for all singularities in the two-jet region and 
can be integrated over the inclusive (i.e.\ two-jet, three-jet and four-jet) phase 
space, this is not sufficient. The infrared singularities yielded 
by the subtraction term in the two-jet region of the four-parton phase space 
exactly cancel those coming from the double real emission matrix element
integrated over the same two-jet region. Furthermore, the subtraction 
term yields only finite terms when integrated over the four-jet region. However, the subtraction term 
generates single infrared singularities when integrated over the three-jet region of the four-parton phase space 
which need to be treated carefully. 
Indeed,
the singular structure in the three-jet region has to be 
mapped out and subtracted from the subtraction term.
This is done by subtracting the sum of all dipole terms appropriate to 
 single-particle singularities from the four-parton matrix element
subtraction term. According to (\ref{eq:sub1}), these read
\begin{equation}
\d \sigma_{NNLO}^{S,1} = 
 N_{in}{\rm d}\Phi_{4}(p_{1},...,p_{4},Q) \,
\sum_{{\rm pairs~} i,j}\;\sum_{k \neq i,j}{\cal D}_{ijk}\,
|{\cal M}_{3}(p_{l},\tilde{p}_{ij},\tilde{p}_{k})|^{2}\;
{\cal F}^{(3)}_{3}(p_{l},\tilde{p}_{ij},\tilde{p}_{k})\;,
\end{equation}
where $l\neq i,j,k$.
With this, the full NNLO subtraction term becomes
\begin{equation}
{\rm d}\sigma_{NNLO}^{S} = {\rm d}\sigma_{NNLO}^{S,0} - 
{\rm d}\sigma_{NNLO}^{S,1} \;.
\end{equation}

Together with
\begin{equation}
{\rm d}\sigma_{NNLO}^{R} = N_{in}\;{\rm d}\Phi_{4}(p_{1},...,p_{4},Q)\;
|{\cal M}_{4}(p_{1},...p_{4})|^{2} 
{\cal F}_{2}^{(4)}(p_{1},p_{2},p_{3},p_{4})\;,
\end{equation}
we obtain the difference between double real and subtracted matrix element as,
\begin{eqnarray}
{\rm d}\sigma_{NNLO}^{R}- {\rm d}\sigma_{NNLO}^{S}&=&
N_{in}\;{\rm d}\Phi_{4}(p_{1},...,p_{4},Q)\;\times \nonumber \\
&&\Bigg[\,|{\cal M}_{4}(p_{1},...p_{4})|^{2}\;
 \left( 
{\cal F}_{2}^{(4)}(p_{1},p_{2},p_{3},p_{4})
-\frac{1}{2}{\cal F}_{2}^{(2)}(\tilde{p}_{134},\tilde{p}_{2}) 
-\frac{1}{2}{\cal F}_{2}^{(2)}(\tilde{p}_{234},\tilde{p}_{1}) 
\right)\nonumber\\
&& + 
\sum_{{\rm pairs~} i,j}\;\sum_{k \neq i,j}{\cal D}_{ijk}\,
|{\cal M}_{3}(p_{l},\tilde{p}_{ij},\tilde{p}_{k})|^{2}\;
{\cal F}^{(3)}_{3}(p_{l},\tilde{p}_{ij},\tilde{p}_{k})\Bigg]\;,
\label{eq:fourpart}
\end{eqnarray}
which denotes the full contribution from the 
four-parton channel to the two-jet production cross section. It is finite 
and can therefore be integrated numerically.

The subtraction term for the one-loop virtual contribution
is constructed in complete 
analogy to the NLO three-particle subtraction term (\ref{eq:NLOirsub})
by using the full one-loop three-particle matrix element. We have,
\begin{eqnarray}
{\rm d}\sigma_{NNLO}^{V,1}-{\rm d}\sigma_{NNLO}^{VS,1} &=&
N_{in}\;{\rm d}\Phi_{3}(p_{1},...,p_{3},Q)\;
|{\cal M}^{V,1}_{3}(p_{1},...p_{3})|^{2}\;\times \nonumber \\
& &\left( 
{\cal F}_{2}^{(3)}(p_{1},p_{2},p_{3})
-\frac{1}{2}{\cal F}_{2}^{(2)}(\widetilde{p}_{13},\tilde{p}_{2}) 
-\frac{1}{2}{\cal F}_{2}^{(2)}(\widetilde{p}_{23},\tilde{p}_{1}) \right).
\end{eqnarray}
The difference between the one-loop virtual contribution and its subtraction term,
${\rm d}\sigma_{NNLO}^{V,1} - {\rm d}\sigma_{NNLO}^{VS,1}$,  is integrated over 
the two-jet and three-jet regions of the three-parton final state phase
space. In the two-jet region, it is finite (in fact, yields a zero
contribution) and 
can be integrated numerically in a 
straightforward manner.
 Although  ${\rm d}\sigma_{NNLO}^{VS,1}$ will not produce 
infrared poles from real emission in the three-jet region 
of the final 
state phase space, it will still generate infrared poles from this 
region, since $|{\cal M}^{V,1}_{3}(p_{1},...p_{3})|^{2}$  contains 
explicit infrared poles from the loop integration. These poles are 
proportional to the tree level three-particle matrix element squared, and 
are canceled by extracting the infrared poles from the real emission 
subtraction term ${\rm d}\sigma_{NNLO}^{S,1} $. To see this, 
we express
\begin{eqnarray}
\d \sigma_{NNLO}^{S,1} &=& 
 N_{in}{\rm d}\Phi_{4}(p_{1},...,p_{4},Q) \,
\sum_{{\rm pairs~} i,j}\;\sum_{k \neq i,j}{\cal D}_{ijk}\,
|{\cal M}_{3}(p_{l},\tilde{p}_{ij},\tilde{p}_{k})|^{2}\;
{\cal F}^{(3)}_{3}(p_{l},\tilde{p}_{ij},\tilde{p}_{k})\; \nonumber\\
&=&  N_{in}\,
\sum_{{\rm pairs~} i,j}\;\sum_{k \neq i,j}
{\rm d}\Phi_{3}(p_{l},\tilde{p}_{ij},\tilde{p}_{k},Q) \,
|{\cal M}_{3}(p_{l},\tilde{p}_{ij},\tilde{p}_{k})|^{2}\;
{\cal F}^{(3)}_{3}(p_{l},\tilde{p}_{ij},\tilde{p}_{k})
\int_{\d\Phi_D} {\cal D}_{ijk}\;.
\end{eqnarray}
Since the sum of 
integrals of the dipole terms over the dipole phase space 
exactly cancels the infrared singularities present in the corresponding 
one-loop matrix element, we find that 
\begin{eqnarray}
 - \d \sigma_{NNLO}^{VS,1} -\d \sigma_{NNLO}^{S,1} &=&
N_{in}{\rm d}\Phi_{3}(p_{1},p_2,p_3,Q) \, \times
\nonumber \\
&& \Bigg[ |{\cal M}^{V,1}_{3}(p_{1},...p_{3})|^{2} \left(
-\frac{1}{2}{\cal F}_{2}^{(2)}(\widetilde{p}_{13},\tilde{p}_{2}) 
-\frac{1}{2}{\cal F}_{2}^{(2)}(\widetilde{p}_{23},\tilde{p}_{1}) 
\right) \nonumber \\
&& \hspace{4mm}
- |{\cal M}_{3}(p_1,p_2,p_3)|^{2}\;
{\cal F}^{(3)}_{3}(p_1,p_2,p_3)
\sum_{{\rm pairs~} i,j; k \neq i,j}
\int_{\d\Phi_D} {\cal D}_{ijk}\, \Bigg] 
\end{eqnarray}
is finite if integrated over the three-jet region of the 
three-parton final state phase 
space. The full contribution from the three-parton channel 
to two-jet final states is therefore
\begin{eqnarray}
\lefteqn{ \d \sigma_{NNLO}^{V,1} - \d \sigma_{NNLO}^{VS,1} 
-\d \sigma_{NNLO}^{S,1} =}\nonumber \\
&& N_{in}{\rm d}\Phi_{3}(p_1,p_2,p_3,Q) \;\Bigg[ 
|{\cal M}^{V,1}_{3}(p_{1},...p_{3})|^{2} \left(
{\cal F}_{2}^{(3)}(p_{1},p_{2},p_{3})
-\frac{1}{2}{\cal F}_{2}^{(2)}(\widetilde{p}_{13},\tilde{p}_{2}) 
-\frac{1}{2}{\cal F}_{2}^{(2)}(\widetilde{p}_{23},\tilde{p}_{1}) 
\right) \nonumber \\
&& \hspace{4mm}
- |{\cal M}_{3}(p_1,p_2,p_3)|^{2}\;
{\cal F}^{(3)}_{3}(p_1,p_2,p_3)
\sum_{{\rm pairs~} i,j; k \neq i,j}
\int_{\d\Phi_D} {\cal D}_{ijk}\, \Bigg]\;, 
\label{eq:threepart}
\end{eqnarray}
which is finite over the full three-parton phase space and can be integrated 
numerically. 

Finally, the subtraction terms $\d \sigma_{NNLO}^{S,0}$ and 
$\d \sigma_{NNLO}^{VS,1}$ integrated over the tripole 
and dipole phase space
cancel the infrared singularities 
present in the two-loop virtual contribution
$\d \sigma_{NNLO}^{V,1}$. Using ${\cal F}_{2}^{(2)}=1$, we find
\begin{eqnarray}
\d \sigma_{NNLO}^{S,0} &=& N_{in}\;{\rm d}\Phi_{4}\;
|{\cal M}_{4}|^{2} 
=  N_{in}\;{\rm d}\Phi_{2} |{\cal M}_{2}|^{2} \int_{\d \Phi_T}
|M_{4}|^{2} \;,\\
\d \sigma_{NNLO}^{VS,1} &=& N_{in}\;{\rm d}\Phi_{4}\;
|{\cal M}^{V,1}_{3}|^{2} 
=  N_{in}\;{\rm d}\Phi_{2} |{\cal M}_{2}|^{2} \int_{\d \Phi_D}
|M^{V,1}_{3}|^{2} \;.
\end{eqnarray}
Adding these to the two-loop virtual contributions,
\begin{eqnarray}
\d \sigma_{NNLO}^{V,2} +
\d \sigma_{NNLO}^{S,0} +
\d \sigma_{NNLO}^{VS,1} &=&  N_{in}\;{\rm d}\Phi_{2} |{\cal M}_{2}|^{2}
\left[ |M^{V,2}_{2}|^{2} +  \int_{\d \Phi_T}
|M_{4}|^{2} +  \int_{\d \Phi_D}
|M^{V,1}_{3}|^{2} \right]\;,
\label{eq:twopart}
\end{eqnarray}
we find an infrared finite result for the two-parton channel.

To summarize schematically,
 the NNLO corrections to the two-jet rate take the following 
structure
\begin{eqnarray}
{\rm d}\sigma_{NNLO}&=&
\hspace{3.3mm}\left [\d \sigma_{NNLO}^{R} - \d \sigma_{NNLO}^{S,0} +
\d \sigma_{NNLO}^{S,1}
\right] \nonumber \\
&&+ \left [\d \sigma_{NNLO}^{V,1} - \d \sigma_{NNLO}^{VS,1} -
\d \sigma_{NNLO}^{S,1}
\right] \nonumber \\
&&+ \left [\d \sigma_{NNLO}^{V,2} + \d \sigma_{NNLO}^{S,0} +
\d \sigma_{NNLO}^{VS,1}
\right] \\
&=& \hspace{2.7mm} {\rm d}\Phi_{4}\;\left[|{\cal M}_{4}|^{2}\,
\left({\cal F}_{2}^{(4)}-{\cal F}_{2}^{(2)}\right) + 
\sum_{ijk} |{\cal M}_{3}|^{2}\, D_{ijk} {\cal F}_{3}^{(3)} \right] \nonumber \\
&& 
+{\rm d}\Phi_{3}\; \left[ 
|{\cal M}_{3}^{V,1}|^2\,\left({\cal F}_{2}^{(3)}
- {\cal F}_{2}^{(2)} \right) 
- \sum_{ijk} |{\cal M}_{3}|^{2}\, \left(\int_{\d \Phi_D} D_{ijk}\right)
 {\cal F}_{3}^{(3)} \right] 
\nonumber \\
& & +{\rm d}\Phi_{2}\;|{\cal M}_{2}|^2 \left[ 
|M^{V,2}_2|^2  + \int_{\d \Phi_T} |M_4|^2 + \int_{\d \Phi_D} 
|M_3^{V,1}|^2 \right]  \;{\cal F}_{2}^{(2)}\;.
\end{eqnarray}
The two-, three- and four-parton final state contributions are made explicit
in this formula as those contain  respectively the corresponding 
parton phase space ${\rm d}\Phi_{i}$. These correspond to (\ref{eq:twopart}),
(\ref{eq:threepart}) and (\ref{eq:fourpart}), 
which are all separately finite and can be integrated 
numerically over the two-, three- and four-parton final space respectively. 

The interpretation of the individual terms is as follows:
the infrared singularities of the four-parton contribution 
$\d \sigma_{NNLO}^{R}$ to two-jet final states 
are subtracted 
through the full matrix element, which is integrated over the inclusive 
phase space, $\d \sigma_{NNLO}^{S,0}$. Since this term also contains 
infrared singularities in the three-jet region, a further subtraction 
term $\d \sigma_{NNLO}^{S,1}$
applicable only in this region,  has to be subtracted off 
$\d \sigma_{NNLO}^{S,0}$. Infrared singularities from the 
one-loop virtual correction to the three-parton contribution to two-jet final
states $\d \sigma_{NNLO}^{V,1}$
are 
subtracted through the full one-loop matrix element integrated over the 
inclusive phase space, $\d \sigma_{NNLO}^{VS,1}$. This term yields explicit 
infrared poles in the three-jet region, which cancel with the 
infrared singularities obtained by integrating   $\d \sigma_{NNLO}^{S,1}$
over the single unresolved phase space appropriate to three-jet final states. 
Finally, all infrared poles produced through the two-jet contribution of the 
inclusive phase space integration of the subtraction terms $\d \sigma_{NNLO}^{S,0}$ 
and $\d \sigma_{NNLO}^{VS,1}$
cancel with the poles from the two-loop virtual corrections,
 $\d \sigma_{NNLO}^{V,2}$. 

The construction of the subtraction terms presented here differs in one
essential point from the procedure proposed in~\cite{weinzierl1}:
we first subtract all double and single unresolved singularities 
from the real radiation contribution $\d \sigma_{NNLO}^{R}$ using 
$\d \sigma_{NNLO}^{S,0}$ and afterwards add in the oversubtracted 
single unresolved singularities through $\d \sigma_{NNLO}^{S,1}$. In contrast, 
\cite{weinzierl1} first subtracts all single unresolved 
singularities from the real radiation contribution 
$\d \sigma_{NNLO}^{R}$ and then constructs  
subtraction terms which account only for the double real
emission contributions in the (already singly subtracted) matrix element. 
The double real emission subtraction terms constructed this way are 
more involved than the corresponding terms employed here. In particular, 
the subtraction terms obtained in~\cite{weinzierl1} can not be related to
four-particle matrix elements in a straightforward manner, which renders 
their analytic phase space integration more difficult. As a matter of fact,
the subtraction terms constructed here can be integrated using the 
methods derived in~\cite{ggh}, while the subtraction terms 
of~\cite{weinzierl1} first require the computation of as yet unknown
classes of integrals~\cite{weinzierl3}.
 
On the other hand, let us note that taking  the subtraction term to be the
full  four-parton matrix element squared we clearly do not separate  
singularities arising from individual partonic configurations. Therefore, more 
work is needed for extending the method derived here to final states with more
than two jets, while the subtraction terms  of~\cite{weinzierl1}   readily
generalize to final states with higher jet multiplicity. 

In the next section, we analytically compute the infrared poles  arising in
each parton level contribution discussed here. For simplicity,  we set
$F_{J}^{(m)}=1$ in the following (corresponding to computing the inclusive
cross section) and discard the single-particle  subtraction term $\d
\sigma_{NNLO}^{S,1}$, whose infrared structure is well  known from NLO
calculations~\cite{cs} and whose contributions exactly cancel.

\section{NNLO contributions}
\setcounter{equation}{0}
\label{sec:nnlo}

As outlined in Section~\ref{sec:contributions} above, two-jet final states at
NNLO accuracy are obtained from the production of up to four final state
partons. In this section, we present the contributions from  two, three and
four-parton final states, elaborating the  structure of infrared singularities
for each channel. The integrated partonic terms  derived in this section  
represent the integrals of the subtraction terms presented in Section 3 and
account for all singularities arising from real radiation.

\subsection{Two-parton final states}
The calculation of the virtual two-loop corrections to 
$\gamma^* \to q\bar q$ (also called quark form factor) 
in dimensional regularization with $d=4-2\e$ space-time dimensions
was performed long ago~\cite{vanneerven}.
In the following, we 
only list  the results, and illustrate the application of 
the infrared factorization formula.  Renormalization of ultraviolet divergences is 
performed in the $\overline{{\rm MS}}$ scheme by 
replacing 
the bare coupling $\alpha_0$ with the renormalized coupling 
$\alpha_s\equiv \alpha_s(\mu^2)$,
evaluated at the renormalization scale $\mu^2$
\begin{equation}
\alpha_0\mu_0^{2\e} S_\e = \alpha_s \mu^{2\e}\left[
1- \frac{11 N - 2 N_f}{6\e}\left(\frac{\alpha_s}{2\pi}\right) 
+{\cal O}(\alpha_s^2) \right]\; ,
\end{equation}
where
\begin{displaymath}
S_\e =(4\pi)^\e e^{-\e\gamma}\qquad \mbox{with Euler constant }
\gamma = 0.5772\ldots
\end{displaymath}
and $\mu_0^2$ is the mass parameter introduced 
in dimensional regularization to maintain a 
dimensionless coupling 
in the bare QCD Lagrangian density.

The renormalized 
amplitude can be written as
\begin{equation}
|{\cal M}\rangle_{q\bar q} = \sqrt{4\pi\alpha}e_q  \left[
|{\cal M}^{(0)}\rangle_{q\bar q} 
+ \left(\frac{\alpha_s}{2\pi}\right) |{\cal M}^{(1)}\rangle_{q\bar q} 
+ \left(\frac{\alpha_s}{2\pi}\right)^2 |{\cal M}^{(2)}\rangle_{q\bar q} 
+ {\cal O}(\alpha_s^3) \right] \;,
\label{eq:renorme}
\end{equation}
where $\alpha$ denotes the electromagnetic coupling constant, 
$e_q$ the quark charge,
$\alpha_s$ the QCD coupling constant at the renormalization scale $\mu$, 
and the $|{\cal M}^{(i)}\rangle$ are the $i$-loop contributions to the 
renormalized amplitude. They are scalars in colour space. 

The squared amplitude, summed over spins, colours and quark flavours, 
is denoted by
\begin{equation}
\langle{\cal M}|{\cal M}\rangle _{q\bar q}
= \sum |{\cal M}(\gamma^* \to q\bar q)|^2 
= {\cal T}_{q\bar q} (q^2)\; .
\end{equation}
The perturbative expansion of ${\cal T}_{q\bar q}
 (q^2)$ at renormalization scale 
$\mu^2 = q^2$ reads:
\begin{eqnarray}
{\cal T}_{q\bar q} &=& 4\pi\alpha\sum_q e_q^2 \Bigg[
{\cal T}_{q\bar q}^{(2)} (q^2) + 
\left(\frac{\alpha_s(q^2)}{2\pi}\right){\cal T}^{(4)}_{q\bar q} (q^2) \nonumber \\
&& \hspace{1.3cm}
+ \left(\frac{\alpha_s(q^2)}{2\pi}\right)^2{\cal T}^{(6)}_{q\bar q} (q^2) 
 + {\cal O}(\alpha_s^3(q^2)) \Bigg] \;,
\end{eqnarray}
where 
\begin{eqnarray}
\label{eq:T2}
{\cal T}^{(2)}_{q\bar q} 
(q^2) &=& \langle{\cal M}^{(0)}|{\cal M}^{(0)}\rangle_{q\bar q}  
= 4 N (1-\e) q^2 \;,\\
\label{eq:T4}
{\cal T}^{(4)}_{q\bar q} (q^2) &=& 
\langle{\cal M}^{(0)}|{\cal M}^{(1)}\rangle_{q\bar q}  +
\langle{\cal M}^{(1)}|{\cal M}^{(0)}\rangle_{q\bar q}  \nonumber \\
& = & \left( N-\frac{1}{N}\right){\cal T}^{(2)}_{q\bar q} (q^2)
  \Bigg[ -\frac{1}{\e^2} - \frac{3}{2\e} - 4 + 
\frac{7\pi^2}{12} 
+ \left( -8 + \frac{7\pi^2}{8} + \frac{7}{3}\zeta_3 \right)
\e \nonumber \\
&& \hspace{2cm}
+ \left( - 16
          + \frac{7\pi^2}{3}
          + \frac{7}{2}\zeta_3
          - \frac{73\pi^4}{1440} \right) \e^2  + {\cal O}(\e^3) \Bigg] \; ,\\
{\cal T}^{(6)}_{q\bar q} (q^2) &=& 
\langle{\cal M}^{(1)}|{\cal M}^{(1)}\rangle_{q\bar q}  +
\langle{\cal M}^{(0)}|{\cal M}^{(2)}\rangle_{q\bar q}  +
\langle{\cal M}^{(2)}|{\cal M}^{(0)}\rangle_{q\bar q}  \;,
\label{eq:T6}
\end{eqnarray}
where $N$ is the number of colours.

In the following, 
we decompose 
 ${\cal T}^{(6)}_{q\bar q} (q^2)$ 
into the contributions from the interference 
of two-loop and tree diagrams,
\begin{eqnarray}
{\cal T}_{q\bar q}^{(6,[2\times 0])} (q^2) &=& 
\langle{\cal M}^{(0)}|{\cal M}^{(2)}\rangle_{q\bar q}  +
\langle{\cal M}^{(2)}|{\cal M}^{(0)}\rangle_{q\bar q}  \\
&=&
\left( N-\frac{1}{N}\right)\;{\cal T}^{(2)}_{q\bar q} (q^2)\;
  \Bigg\{ \; N \Bigg[ \frac{1}{4\e^4} + \frac{17}{8\e^3}
+ \frac{1}{\e^2} \left( \frac{433}{144} -\frac{\pi^2}{2} \right)
\nonumber \\&&
+ \frac{1}{\e} \left( \frac{4045}{864} -\frac{83\pi^2}{48} 
+\frac{7}{12}\zeta_3 \right)
+\left(-\frac{9083}{5184}
          - \frac{2153\pi^2}{864}
          + \frac{13}{9}\zeta_3           + \frac{263\pi^4}{1440}
   \right) \Bigg]\nonumber \\
&&+\frac{1}{N} \Bigg[ -\frac{1}{4\e^4} - \frac{3}{4\e^3}
+ \frac{1}{\e^2} \left( -\frac{41}{16} +\frac{13\pi^2}{24} \right)
\nonumber \\&&
+ \frac{1}{\e} \left( -\frac{221}{32} +\frac{3\pi^2}{2} 
+\frac{8}{3}\zeta_3 \right)
+\left(-\frac{1151}{64}
          + \frac{475\pi^2}{96}
          + \frac{29}{4}\zeta_3   - \frac{59\pi^4}{288}
   \right) \Bigg]\nonumber \\
&&+ N_f \Bigg[ -\frac{1}{4\e^3} - \frac{1}{9\e^2}
+ \frac{1}{\e} \left( \frac{65}{216} +\frac{\pi^2}{24} \right)
+ \left( \frac{4085}{1296} -\frac{91\pi^2}{216} 
+\frac{1}{18}\zeta_3 \right)  + {\cal O}(\e) 
\Bigg]
 \Bigg\}\;,
\end{eqnarray}
as well as the one-loop self-interference, 
\begin{eqnarray}
{\cal T}_{q\bar q}^{(6,[1\times 1])} (q^2) &=& 
\langle{\cal M}^{(1)}|{\cal M}^{(1)}\rangle_{q\bar q}  \nonumber \\
&=&
{\cal T}^{(2)}_{q\bar q}(q^2)\;\left( N-\frac{1}{N}\right)^2\;
  \Bigg[ \frac{1}{4\e^4} + \frac{3}{4\e^3}
+ \frac{1}{\e^2} \left( \frac{41}{16} -\frac{\pi^2}{24} \right)
\nonumber \\&&
+ \frac{1}{\e} \left( 7 -\frac{\pi^2}{8} 
+\frac{7}{6}\zeta_3 \right)
+\left(18
          - \frac{41\pi^2}{96}
          - \frac{7}{2}\zeta_3           - \frac{7\pi^4}{480}
   \right) + {\cal O}(\e)  \Bigg]
\end{eqnarray}

The pole structure of ${\cal T}_{q\bar q}^{(4)}$ and of  
both
contributions to ${\cal T}_{q\bar q}^{(6)}$ 
can be expressed using an infrared factorization formula, obtained from
resummation~\cite{catani}.
To this end, we write
\begin{eqnarray}
{\cal T}_{q\bar q}^{(4)}(q^2) &=&\Poles_{q\bar q}^{(1 \times 0)}(q^2)
+\Finite_{q\bar q}^{(1 \times 0)}(q^2), \\
{\cal T}_{q\bar q}^{(6, [i\times j])}(q^2)
 &=& \Poles_{q\bar q}^{(i \times j)}(q^2)
+\Finite_{q\bar q}^{(i \times j)}(q^2).
\end{eqnarray} 
$\Poles_{q\bar q}$ contains infrared singularities that will be  analytically
canceled by those occurring in radiative processes of the
same order.
$\Finite$ is the renormalized remainder, which is finite as $\epsilon \to 0$.
For simplicity we set the renormalization scale $\mu^2 = q^2$.

The infrared factorization formula shows how to predict 
infrared pole structure of the one-loop and 
two-loop contributions renormalized in the 
\MSbar\ scheme in terms of the tree and renormalized one-loop amplitudes,
$\ket{\cm^{(0)}}$ and $\ket{\cm^{(1)}}$ respectively. At one-loop
it yields  
\begin{equation}
\Poles_{q\bar q}^{(1 \times 0)} = 2 \Re 
\bra{\cm^{(0)}} {\bom I}^{(1)}(\epsilon) \ket{\cm^{(0)}}, 
\label{eq:polesx}
\end{equation}
while at two loops
\begin{eqnarray}
\label{eq:polesa}
\lefteqn{\hspace{-2cm}\Poles_{q\bar q}^{(2 \times 0)} = 
2 \Re \Biggl[  -\frac{1}{2}\bra{\cm^{(0)}} {\bom I}^{(1)}(\epsilon){\bom
I}^{(1)}(\epsilon) \ket{\cm^{(0)}}
  -\frac{\beta_0}{\epsilon}  
\,\bra{\cm^{(0)}} {\bom I}^{(1)}(\epsilon) 
\ket{\cm^{(0)}}}
 \nonumber\\
&& 
+\,  \bra{\cm^{(0)}} {\bom I}^{(1)}(\epsilon)  \ket{\cm^{(1)}}
 \nonumber\\
&& 
+
e^{-\epsilon \gamma } \frac{ \Gamma(1-2\epsilon)}{\Gamma(1-\epsilon)} 
\left(\frac{\beta_0}{\epsilon} + K\right)
\bra{\cm^{(0)}} {\bom I}^{(1)}(2\epsilon) \ket{\cm^{(0)}}\nonumber \\
&&+ \, \bra{\cm^{(0)}}{\bom H}^{(2)}(\epsilon)\ket{\cm^{(0)}} \Biggr],
\end{eqnarray}
and
\begin{equation}
\label{eq:polesb}
\Poles_{q\bar q}^{(1 \times 1)} = 
\Re \Biggl[  2 \bra{\cm^{(1)}} {\bom I}^{(1)}(\epsilon)  \ket{\cm^{(0)}}
-\bra{\cm^{(0)}} {\bom I}^{(1)\dagger}(\epsilon)
{\bom I}^{(1)}(\epsilon) \ket{\cm^{(0)}}
\, \Biggr],
\end{equation}
where the constant $K$ is
\begin{equation}
K = \left( \frac{67}{18} - \frac{\pi^2}{6} \right) \CA - 
\frac{10}{9} T_R \NF.
\end{equation}
It should be noted that, in this prescription, part of the finite 
terms in ${\cal T}_{q\bar q}^{(4)}$  and
${\cal T}_{q\bar q}^{(6,[i\times j])}$ are accounted for 
by the ${\cal O}(\e^0)$
expansion of $\Poles_{q\bar q}^{(i \times j)}$.

For the process under consideration, 
there is only one colour structure present at
tree level which is just the contraction of 
the quark
and antiquark
colours $i$ and $j$, it is simply $\delta_{ij}$. Adding higher loops does not
introduce additional colour structures, and the amplitudes are therefore
scalars.
Similarly, 
the infrared singularity operator 
$\bom{I}^{(1)}(\epsilon)$ is a $1 \times 1$ matrix in the colour space. 
By explicit evaluation of $\Poles_{q\bar q}^{(i \times j)}$, one can show that 
$\bom{I}^{(1)}(\epsilon)$ is only determined up to a finite constant
(which we shall label $c_1$ below) that 
can be chosen arbitrarily. In fact, in~\cite{cs}, where 
$\bom{I}^{(1)}(\epsilon)$ was derived for the first time, a constant term 
arising from the finite contributions to the dipole matrix elements 
was found. This term was omitted in~\cite{catani}.  

Including  $c_1$, $\bom{I}^{(1)}(\epsilon)$ 
is given by,
\begin{equation}
\bom{I}^{(1)}(\epsilon)
=
- \frac{e^{\epsilon\gamma}}{2\Gamma(1-\epsilon)} \Biggl[
\frac{N^2-1}{2N}
\left(\frac{2}{\epsilon^2}+\frac{3}{\epsilon}+c_1\right)
{\tt S}_{12}\Biggr ]\; ,\label{eq:I1}
\end{equation}
(since we have set $\mu^2 = q^2$):
\begin{equation}
{\tt S}_{12} = \left(-\frac{\mu^2}{q^2}\right)^{\epsilon}
= \left(-1\right)^{\epsilon}.
\end{equation}
Note that on expanding ${\tt S}_{12}$,
imaginary parts are generated, 
the sign of which is fixed by the small imaginary
part $+i0$ of $q^2$.
Other combinations such as 
$\bra{\cm^{(0)}}\bom{I}^{(1)\dagger}(\epsilon)$  are obtained by using the hermitian conjugate
operator $\bom{I}^{(1)\dagger}(\epsilon)$, 
where the only practical change is that the sign of the
imaginary part of ${\tt S}_{12}$ is reversed.

Finally, the last term of Eq.~(\ref{eq:polesa}) that involves 
${\bom H}^{(2)}(\epsilon)$ 
produces only a single pole in $\epsilon$ and is given by, 
\begin{equation}
\label{eq:htwo}
\bra{\cm^{(0)}}{\bom H}^{(2)}(\epsilon)\ket{\cm^{(0)}} 
=\frac{e^{\epsilon \gamma}}{4\,\epsilon\,\Gamma(1-\epsilon)} H^{(2)} 
\braket{\cm^{(0)}}{\cm^{(0)}} \;,  
\end{equation}
where the constant $H^{(2)}$ is renormalization-scheme-dependent, but 
independent of the choice of $c_1$.
As with the single pole parts of $\bom{I}^{(1)}(\epsilon)$,
the process-dependent
$H^{(2)}$ can be constructed by counting the number of
radiating partons present in the event.
In our case, there is only a 
 quark--antiquark pair present in the final
state, so that 
\begin{equation}
H^{(2)} =  2H^{(2)}_{q}
\end{equation}
where in the \MSbar\ scheme
\begin{eqnarray}
H^{(2)}_q &=&
\left({7\over 4}\zeta_3+{\frac {409}{864}}- {\frac {11\pi^2}{96}}
\right)N^2
+\left(-{1\over 4}\zeta_3-{41\over 108}-{\pi^2\over 96}\right)
+\left(-{3\over 2}\zeta_3-{3\over 32}+{\pi^2\over 8}\right){1\over
N^2}\nonumber \\
&&
+\left({\pi^2\over 48}-{25\over 216}\right){(N^2-1)N_F\over N}\;.
\end{eqnarray}

As a result, one finds the finite parts 
\begin{eqnarray}
\frac{\Finite^{(1 \times 0)}_{q\bar q}(q^2)}{4Nq^2} 
&=& \left(N-\frac{1}{N}\right) \, \left(-4 + \frac{c_1}{2}\right)
 \\
\frac{\Finite^{(2 \times 0)}_{q\bar q}(q^2)}{4Nq^2} 
&=& \left(N-\frac{1}{N}\right)
\Bigg[     N \left(
          - \frac{81659}{5184}
          - \frac{5\pi^2}{36}
          + \frac{389}{72}\,\zeta_3
          + \frac{49\pi^4}{1440}
          + c_1\left(\frac{31}{36} - \frac{\pi^2}{12} + \frac{c_1}{16}\right)
          \right)\nonumber \\
&&       +  \frac{1}{N}\left(
          - \frac{255}{64}
          - \frac{29\pi^2}{48}
          + \frac{15}{4} \,\zeta_3
          + \frac{11\pi^4}{360}
          + c_1\left(1 - \frac{c_1}{16}\right)
          \right)\nonumber \\
&&       +  N_f \left(
            \frac{4085}{1296}
          + \frac{7\pi^2}{72}
          - \frac{1}{36}\,\zeta_3
          - \frac{5c_1}{18}\right)\Bigg], \\
\frac{\Finite^{(1 \times 1)}_{q \bar q}(q^2)}{4Nq^2} &=& 4\, 
\left(N-\frac{1}{N}\right)^2 \, \frac{(8-c_1)^2}{64}.
\end{eqnarray}

\subsection{Three-parton final states}

The three-particle final state yielding two jets is $\gamma^* \to q\bar q g$ 
in the configuration where only two jets are formed from three partons. 

The renormalized 
amplitude for this process can be written as
\begin{equation}
|{\cal M}\rangle_{q\bar q g} = \sqrt{4\pi\alpha}e_q\, \sqrt{4\pi\alpha_s}\,
 \left[|{\cal M}^{(0)}\rangle_{q\bar q g} 
+ \left(\frac{\alpha_s}{2\pi}\right) |{\cal M}^{(1)}\rangle_{q\bar q g} 
+ {\cal O}(\alpha_s^2) \right] \;,
\end{equation}
where $|{\cal M}^{(i)}\rangle$ are the $i$-loop contributions to the 
renormalized amplitude. 
The squared amplitude, summed over spins, colours and quark flavours, 
is denoted by
\begin{equation}
\langle{\cal M}|{\cal M}\rangle_{q\bar q g}
 = \sum |{\cal M}(\gamma^* \to q\bar q g)|^2. 
\end{equation}
The perturbative expansion of the squared amplitude at renormalization scale 
$\mu^2 = q^2$ reads:
\begin{eqnarray}
\langle{\cal M}|{\cal M}\rangle_{q\bar q g}
&=& 4\pi\alpha\sum_q e_q^2\,8\pi^2 
\Bigg[\left(\frac{\alpha_s(q^2)}{2\pi}\right)
\langle{\cal M}^{(0)}|{\cal M}^{(0)}\rangle_{q\bar q g} 
\nonumber \\
&& \hspace{1.3cm}
+ \left(\frac{\alpha_s(q^2)}{2\pi}\right)^2
\left(\langle{\cal M}^{(0)}|{\cal M}^{(1)}\rangle_{q\bar q g} 
+\langle{\cal M}^{(1)}|{\cal M}^{(0)}\rangle_{q\bar q g} \right)
+ {\cal O}(\alpha_s^3(q^2)) \Bigg]. 
\end{eqnarray}

The squared amplitudes are integrated over the three-parton phase space and multiplied 
by ${\cal F}_{3}^{(2)}$, the jet function ensuring that out of three partons 
a two-jet final state is formed.
The two-jet final state phase space for these three-parton 
configurations contains the regions where the gluon is 
either collinear or soft, and subtraction terms have to be introduced 
to extract the resulting infrared singularities from the 
partonic cross sections.
As outlined in Section~\ref{sec:irsub}, these subtraction terms
are taken to be the 
full squared amplitudes associated with $\gamma^{*}  \to q\bar q g $ up to 
the one-loop order.
Those are integrated over the three-parton phase space factorized in a dipole 
phase space factor ${\rm d}\Phi_{D}$ and the integrated two-parton phase 
space, $P_{2}$.
The momenta associated with this dipole phase space are the two dipole momenta 
(emitter and spectator) and one unresolved (collinear or soft gluon) .
The infrared poles present at ${\cal O}(\alpha_{s})$ are made 
explicit as follows,  
\begin{eqnarray}
{\cal T}^{(4)}_{q\bar q g} (q^2) &=& 8\pi^2 \int \d \Phi_D 
\langle{\cal M}^{(0)}|{\cal M}^{(0)}\rangle_{q\bar q g} \\
&=& 
\left( N-\frac{1}{N}\right)\;{\cal T}^{(2)}_{q\bar q} (q^2)\;
  \Bigg[
\frac{1}{\e^2} + \frac{3}{2\e} + \frac{19}{4} - 
\frac{7\pi^2}{12} 
+ \left( \frac{109}{8} - \frac{7\pi^2}{8} - \frac{25}{3}\zeta_3 \right)
\e \nonumber \\
&& \hspace{2cm}
+ \left( \frac{639}{16}
          - \frac{133\pi^2}{48}
          - \frac{25}{2}\zeta_3
          - \frac{71\pi^4}{1440} \right) \e^2  + {\cal O}(\e^3) \Bigg] \; .
\end{eqnarray} 
The infrared poles in ${\cal T}^{(4)}_{q\bar q g}$ cancel all infrared poles 
present in the one-loop two-particle final state 
contribution ${\cal T}^{(4)}_{q\bar q}$ in (\ref{eq:polesx}):
\begin{equation}
{\cal T}^{(4)}_{q\bar q g} = \Poles^{(0\times 0)}_{q\bar q g} 
+ \Finite^{(0\times 0)}_{q\bar q g}\;,
\end{equation}
 with 
\begin{eqnarray}
\Poles^{(0\times 0)}_{q\bar q g} &=& - 2 \Re 
\bra{\cm^{(0)}} {\bom I}^{(1)}(\epsilon) \ket{\cm^{(0)}} ,
\\
\frac{\Finite^{(0\times 0)}_{q\bar q g}}{4Nq^2}
 &=& \left(N-\frac{1}{N}\right) \, 
\left(\frac{19}{4}-\frac{c_1}{2}\right)
\;.
\end{eqnarray}

At ${\cal O}(\alpha_{s}^2)$, the renormalized matrix element is 
$\langle{\cal M}^{(0)}|
{\cal M}^{(1)}\rangle_{q\bar q g} +
\langle{\cal M}^{(1)}|{\cal M}^{(0)}\rangle_{q\bar q g}$.
To cancel the poles present in the single unresolved contributions 
related to this virtual matrix element,
a subtraction term is required, which is chosen to be 
the virtual matrix element itself. The analytic integration of this 
subtraction term over the dipole phase space is
obtained by first reducing all terms in the integral to three master
integrals, solving iteratively integration-by-parts identities as outlined 
in~\cite{laporta,gr} for two-loop integrals and in~\cite{ggh} for 
four-particle phase space integrals. For this step, extensive use of the 
algebraic programming language {\tt FORM}~\cite{form} is made. 
As a result, one obtains a linear
combination of  three master integrals, 
denoted by $V_{5,a}, V_{5,b}$ and $V_{8}$, which are listed in the appendix.
Inserting these, one obtains
\begin{eqnarray}
{\cal T}^{(6)}_{q\bar q g} (q^2) &=& 8\pi^2 \int \d \Phi_D\left( 
\langle{\cal M}^{(0)}|{\cal M}^{(1)}\rangle_{q\bar q g} + 
\langle{\cal M}^{(1)}|{\cal M}^{(0)}\rangle_{q\bar q g}\right) \\
&=& 
\left( N-\frac{1}{N}\right)\;{\cal T}^{(2)}_{q\bar q} (q^2)\;
  \Bigg\{ 
\; N \Bigg[ -\frac{5}{4\e^4} - \frac{67}{12\e^3}
+ \frac{1}{\e^2} \left( -\frac{141}{8} +\frac{13\pi^2}{8} \right)
\nonumber \\&&
+ \frac{1}{\e} \left( -\frac{1481}{24} +\frac{107\pi^2}{18} 
+\frac{55}{3}\zeta_3 \right)
+\left(-\frac{10385}{48}
          + \frac{64\pi^2}{3}
          + \frac{1265}{18}\zeta_3           - \frac{41\pi^4}{96}
   \right) \Bigg]\nonumber \\
&&+\frac{1}{N} \Bigg[ \frac{1}{\e^4} + \frac{3}{\e^3}
+ \frac{1}{\e^2} \left( \frac{93}{8} -\frac{4\pi^2}{3} \right)
\nonumber \\&&
+ \frac{1}{\e} \left( \frac{79}{2} -\frac{15\pi^2}{4} 
-\frac{53}{3}\zeta_3 \right)
+\left(\frac{1069}{8}
          - \frac{697\pi^2}{48}
          - \frac{91}{2}\zeta_3   + \frac{19\pi^4}{72}
   \right) \Bigg]\nonumber \\
&&+ N_f \Bigg[ \frac{1}{3\e^3} + \frac{1}{2\e^2}
+ \frac{1}{\e} \left( \frac{19}{2} -\frac{7\pi^2}{36} \right)
+ \left( \frac{109}{24} -\frac{7\pi^2}{24} 
-\frac{25}{9}\zeta_3 \right)  
\Bigg]+ {\cal O}(\e) 
 \Bigg\}\; .
\end{eqnarray}

The infrared poles in ${\cal T}^{(6)}_{q\bar q g}$ cancel part of the 
infrared poles from the two-particle final state 
contribution ${\cal T}^{(6)}_{q\bar q}$. This cancellation becomes
evident if we decompose
\begin{equation}
{\cal T}^{(6)}_{q\bar q g} = \Poles^{(1\times 0)}_{q\bar q g} 
+ \Finite^{(1\times 0)}_{q\bar q g},
\end{equation}
and identify individual terms of $\Poles_{q\bar q g}$ with terms from 
the infrared factorization formulae (\ref{eq:polesa}) and (\ref{eq:polesb}),
\begin{eqnarray}
 \Poles^{(1\times 0)}_{q\bar q g} &=& 2 {\cal R} \Bigg[
-  \bra{\cm^{(0)}} {\bom I}^{(1)}(\epsilon)  \ket{\cm^{(1)}}
  + \frac{\beta_0}{\epsilon}  
\,\bra{\cm^{(0)}} {\bom I}^{(1)}(\epsilon) 
\ket{\cm^{(0)}} - \bra{\cm^{(1)}} {\bom I}^{(1)}(\epsilon)  \ket{\cm^{(0)}}
\nonumber \\
&& \hspace{8mm}- \bra{\cm^{(0)}}{\bom H}_V^{(2)}(\epsilon)\ket{\cm^{(0)}} + 
\frac{1}{2}\,
\bra{\cm^{(0)}}{\bom S}_V^{(2)}(\epsilon)\ket{\cm^{(0)}}
\Bigg]\; .
\label{eq:polesc}
\end{eqnarray}
The first two terms in this formula cancel with terms in 
 (\ref{eq:polesa}), while the third term cancels with a term in 
(\ref{eq:polesb}). The last two terms contain the remaining poles, and will 
be discussed in detail now.

 For the process under consideration, the 
infrared singularity operator ${\bom I}^{(1)}$ (\ref{eq:I1}) contains 
only a single colour structure $(N^2-1)/(2N) = C_F$, as does the one-loop 
squared matrix element $T_{q\bar q}^{(4)}$. By inserting 
 ${\bom I}^{(1)}$ (\ref{eq:I1}) in (\ref{eq:polesc}), one finds that 
all $1/\e^4$ and $1/\e^3$ poles 
of ${\cal T}^{(6)}_{q\bar q g}$ which are proportional to the colour 
structures 
 $[(N^2-1)/(2N)]^2=C_F^2$ 
and $(N^2-1)/(2N)N_f=C_F N_f$ are correctly accounted for. This 
cancellation is independent of the choice of the constant term $c_1$ in 
 ${\bom I}^{(1)}$. The $1/\e^2$ terms proportional to these colour structures
in (\ref{eq:polesc}) depend on $c_1$. They match the corresponding terms 
in  ${\cal T}^{(6)}_{q\bar q g}$ if we choose
\begin{equation}
c_1 = \frac{43}{4} - \frac{\pi^2}{3}\; .
\label{eq:c1}
\end{equation}
With this choice, the first three terms of   (\ref{eq:polesc}) correctly 
account for all $1/\e^4$ to  $1/\e^2$ poles proportional to 
 $[(N^2-1)/(2N)]^2=C_F^2$ 
and $(N^2-1)/(2N)N_f=C_F N_f$ in ${\cal T}^{(6)}_{q\bar q g}$, and 
all remaining single pole terms in these two colour structures can be 
attributed to the ${\bom H}_V^{(2)}(\epsilon)$ term in  (\ref{eq:polesc}).

Pole terms corresponding to the colour structure $N(N^2-1)/(2N)=C_FC_A$ can be 
generated only through the combination of ${\bom I}^{(1)}$ and $\beta_0/\e$ 
in (\ref{eq:polesc}), thus yielding at most $1/\e^3$ poles. In 
${\cal T}^{(6)}_{q\bar q g}$, one does however observe $1/\e^4$ terms 
in this colour structure, as first pointed out in~\cite{leshouches}. 
These terms have no analogue in the two-particle final state 
contributions, and must thus cancel among three and four-particle final 
states. We observe that these terms are proportional to the one-loop 
soft gluon current squared matrix element derived in~\cite{cg} (which is a scalar in colour 
space for this particular process), 
integrated over the dipole phase space.  Explicitly, we find,
\begin{equation}
\bra{\cm^{(0)}}{\bom S}_V^{(2)}(\epsilon)\ket{\cm^{(0)}}
 = S_V^{(2)}\, \braket{\cm^{(0)}}{\cm^{(0)}}\; ,
\end{equation}
with
\begin{eqnarray}
 S_V^{(2)} &=& - \frac{e^{2\e \gamma}}{1+\e}
\left[(N^2-1)\, \frac{1}{\e^2}\, \frac{\Gamma^4(1-\e)\Gamma^3(1+\e)}
{\Gamma^2(1-2\e)\Gamma(1+2\e)} \right]\,
\int \d \Phi_D \left(\frac{q^2}{s_{13}s_{23}}\right)^{1+\e}\\
&=& \left(N^2-1\right) \, \Bigg[ -\frac{1}{4\e^4} - \frac{3}{4\e^3}
+ \frac{1}{\e^2}\,\left(-\frac{13}{4}  + \frac{7\pi^2}{24} \right)
+ \frac{1}{\e}\,\left(-\frac{51}{4}  + \frac{7\pi^2}{8} 
+ \frac{14}{3} \zeta_3  \right)\nonumber \\&& \hspace{1.5cm}
+\left(-\frac{205}{4}  + \frac{91\pi^2}{24} 
+ 14 \zeta_3 +\frac{7\pi^4}{480} \right) + {\cal O}(\e) \Bigg] \;.
\end{eqnarray}
The integrated  soft gluon current accounts for all $1/\e^4$ to  $1/\e^2$
poles proportional to $N(N^2-1)/(2N)=C_FC_A$ in ${\cal T}^{(6)}_{q\bar q g}$, 
the remaining single poles are again attributed to 
${\bom H}_V^{(2)}(\epsilon)$, which finally reads,
\begin{equation}
\bra{\cm^{(0)}}{\bom H}_V^{(2)}(\epsilon)\ket{\cm^{(0)}} 
=\frac{e^{\epsilon \gamma}}{4\,\epsilon\,\Gamma(1-\epsilon)} H_V^{(2)} 
\braket{\cm^{(0)}}{\cm^{(0)}} \;,  
\end{equation}
with
\begin{equation}
H^{(2)}_V = 2 H^{(2)}_{V,q}\;.
\end{equation}
The expression for $H^{(2)}_{V,q}$ in the \MSbar\ scheme is given by, 
\begin{eqnarray}
H^{(2)}_{V,q}&=&
\left(-11\zeta_3+{\frac {409}{24}}+ {\frac {\pi^2}{18}}
\right)N^2
+\left(26\zeta_3-{1655\over 48}-{\pi^2\over 18}\right)
+\left(-15\zeta_3+{279\over 16}\right){1\over
N^2}\nonumber \\
&&
+\left(-{\pi^2\over 18}+{5\over 24}\right){(N^2-1)N_F\over N}\;.
\end{eqnarray}

To summarize the discussion of the infrared pole terms arising from 
three-particle final states at this order, we have demonstrated that 
all double and higher poles can either be identified with terms 
present in the infrared factorization formulae for the virtual corrections
(provided an appropriate finite constant is 
inserted in the infrared singularity operator  ${\bom I}^{(1)}$)
or with the dipole phase space integral of 
the soft gluon current. The single poles are attributed to a constant
$H^{(2)}_V$, whose origin (like the origin of the $H^{(2)}$ appearing 
in the double virtual corrections) is not fully understood at present. 
After subtraction of the poles, one is left with,
\begin{eqnarray}
\frac{\Finite^{(1\times 0)}_{q\bar q g}}{4Nq^2} &=& \left(N-\frac{1}{N}\right)
\Bigg[     N \left(
          - \frac{5549}{48}
          + \frac{41\pi^2}{24}
          + \frac{143}{3}\,\zeta_3
          + \frac{41\pi^4}{180}
          \right)\nonumber \\
&&       +  \frac{1}{N}\left(
           \frac{673}{8}
          - \frac{7\pi^2}{24}
          - \frac{75}{2} \,\zeta_3
          - \frac{73\pi^4}{180}
          \right)
       +  N_f \left(
            \frac{109}{24}
          - \frac{8}{3}\,\zeta_3
          \right)\Bigg]. 
\end{eqnarray}

\subsection{Four-parton final states}
Three different tree level processes can yield 
 four-parton final states relevant to two-jet final states: 
 $\gamma^* \to q\bar q q'\bar q'$ ($q\neq q'$), $\gamma^* \to q\bar q q\bar q$ 
and
$\gamma^* \to q\bar q gg$.  
The 
amplitude for each process can be written as
\begin{equation}
|{\cal M}\rangle_{q\bar q ij} = \sqrt{4\pi\alpha}e_q  \,
4\pi\alpha_s \,
\left[ |{\cal M}^{(0)}\rangle_{q\bar q ij} 
+ {\cal O}(\alpha_s) \right]\;,
\end{equation}
where $ij = q'\bar q', q\bar q, gg$.

The squared amplitude, summed over spins, colours and quark flavours, 
is denoted by
\begin{equation}
\langle{\cal M}|{\cal M}\rangle_{q\bar q ij}
 = \sum |{\cal M}(\gamma^* \to q\bar q ij)|^2 .
\end{equation}
The perturbative expansion of the squared amplitude at renormalization scale 
$\mu^2 = q^2$ reads:
\begin{eqnarray}
\langle{\cal M}|{\cal M}\rangle_{q\bar q ij}
&=& 4\pi\alpha\sum_q e_q^2\, 64\pi^4\,
\Bigg[\left(\frac{\alpha_s(q^2)}{2\pi}\right)^2\,
\langle{\cal M}^{(0)}|{\cal M}^{(0)}\rangle_{q\bar q ij} 
+ {\cal O}(\alpha_s^3(q^2)) \Bigg] .
\end{eqnarray}

These amplitudes need to be integrated over the four-parton phase space and in
order  to give rise to contributions to the two-jet rate, they need to be
multiplied by the  corresponding jet-function ${\cal F}_{4}^{(2)}$. Indeed, two
of the four final state  partons are experimentally unresolved. The two-jet final
state phase space, however also contain regions where one or two particles are
theoretically unresolved; the associated partonic cross sections  contain
single and double singularities.  Those are canceled analytically through the
introduction of subtraction terms.  The appropriate subtraction terms for the 
infrared singular structure of these real-real corrections
are the  corresponding matrix elements themselves as explained in 
Section~\ref{sec:irsub} above. These are to be integrated
over the  tripole phase space factor ${\rm d}\Phi_{T}$ (\ref{eq:triphase}),
which is a factorized form of the four-parton phase space ${\rm d}\Phi_{4}$. These
integrals have been discussed in detail in~\cite{ggh}, where it  was
demonstrated that any integral appearing in this context can be  expressed as a
linear combination of four master integrals  $R_{4}$, $R_{6}$, $R_{8a}$ and
$R_{8b}$, which were derived in  \cite{ggh} and are listed in the appendix for
completeness.
As a result, one finds the integrated subtraction terms,
\begin{eqnarray}
{\cal T}^{(6)}_{q\bar q q' \bar q'} (q^2) &=&
64 \pi^4 \,\int \d \Phi_T  \langle{\cal M}^{(0)}|
{\cal M}^{(0)}\rangle_{q\bar q q'\bar q'}\\
&=& 
{\cal T}^{(2)}_{q\bar q} (q^2)\;\left( N-\frac{1}{N}\right)\; 
(N_f-1)\nonumber 
\\ && 
  \Bigg[ 
- \frac{1}{12\e^3} - \frac{7}{18\e^2}
+ \frac{1}{\e} \left( -\frac{407}{216} +\frac{11\pi^2}{72} \right)
+ \left( - \frac{11753}{1296} +\frac{77\pi^2}{108} 
+\frac{67}{18}\zeta_3 \right)  
+ {\cal O}(\e) \Bigg]\\
{\cal T}^{(6)}_{q\bar q q \bar q} (q^2) &=&
64 \pi^4 \,
\int \d \Phi_T  \langle{\cal M}^{(0)}|{\cal M}^{(0)}\rangle_{q\bar q q\bar q}\\
&=& \frac{1}{N_f-1} {\cal T}^{(6)}_{q\bar q q' \bar q'} (q^2)
+ {\cal T}^{(2)}_{q\bar q} (q^2)\;\left( N-\frac{1}{N}\right)\; \nonumber\\
&& \frac{1}{N} \,\left[ 
 \frac{1}{\e} \left( \frac{13}{16} -\frac{\pi^2}{8} + 
\frac{1}{2}\zeta_3\right)
+ \left(  \frac{339}{32} -\frac{17\pi^2}{24} 
-\frac{21}{4}\zeta_3 +\frac{2\pi^4}{45} \right)  
+ {\cal O}(\e) 
\right] \\
{\cal T}^{(6)}_{q\bar q gg} (q^2) &=&
64 \pi^4 \,
\int \d \Phi_T  \langle{\cal M}^{(0)}|{\cal M}^{(0)}\rangle_{q\bar q gg}\\
&=&
\left( N-\frac{1}{N}\right)\;{\cal T}^{(2)}_{q\bar q} (q^2)\;
  \Bigg\{ \; N 
\Bigg[ \frac{3}{4\e^4} + \frac{65}{24\e^3}
+ \frac{1}{\e^2} \left( \frac{217}{18} -\frac{13\pi^2}{12} \right)
\nonumber \\&&
+ \frac{1}{\e} \left( \frac{43223}{864} -\frac{589\pi^2}{144} 
-\frac{71}{4}\zeta_3 \right)
+\left(\frac{1076717}{5184}
          - \frac{7955\pi^2}{432}
          - \frac{1327}{18}\zeta_3   + \frac{373\pi^4}{1440}
   \right) \Bigg] \nonumber \\
&&+\frac{1}{N} \Bigg[ -\frac{1}{2\e^4} - \frac{3}{2\e^3}
+ \frac{1}{\e^2} \left( -\frac{13}{2} +\frac{3\pi^2}{4} \right)
\nonumber \\&&
+ \frac{1}{\e} \left( -\frac{845}{32} +\frac{9\pi^2}{4} 
+\frac{40}{3}\zeta_3 \right)
+\left(-\frac{6921}{64}
          + \frac{473\pi^2}{48}
          + 40\zeta_3           - \frac{17\pi^4}{144}
   \right) \Bigg]
+ {\cal O}(\e) 
\Bigg]
 \Bigg\}\;.
\end{eqnarray}

The infrared poles from these four-parton final states exactly
cancel the infrared 
poles present in the two- and three-particle final states derived 
above. To see this cancellation, we consider 
\begin{equation}
{\cal T}^{(6)}_{q\bar q q' \bar q'} +
{\cal T}^{(6)}_{q\bar q q \bar q} +
{\cal T}^{(6)}_{q\bar q gg} = \Poles^{(0 \times 0)}_{q\bar q (ij)} 
+ \Finite^{(0 \times 0)}_{q\bar q (ij)}\;,
\end{equation}
and identify
\begin{eqnarray}
\label{eq:polesd}
\Poles^{(0\times 0)}_{q\bar q (ij)} &=&
 \Re \Biggl[  \bra{\cm^{(0)}} {\bom I}^{(1)}(\epsilon){\bom
I}^{(1)}(\epsilon) \ket{\cm^{(0)}}
-2 e^{-\epsilon \gamma } \frac{ \Gamma(1-2\epsilon)}{\Gamma(1-\epsilon)} 
\left(\frac{\beta_0}{\epsilon} + K\right)
\bra{\cm^{(0)}} {\bom I}^{(1)}(2\epsilon) \ket{\cm^{(0)}}\nonumber \\
&&\hspace{7mm}+\bra{\cm^{(0)}} {\bom I}^{(1)\dagger}(\epsilon)
{\bom I}^{(1)}(\epsilon) \ket{\cm^{(0)}}
- 2 \, \bra{\cm^{(0)}}{\bom H}_R^{(2)}(\epsilon)\ket{\cm^{(0)}}\nonumber \\
&& \hspace{7mm}
-  \bra{\cm^{(0)}}{\bom S}_V^{(2)}(\epsilon)\ket{\cm^{(0)}}\Biggr]\;.
\end{eqnarray}

As for the three-parton final states above, we have to fix $c_1$ according to
(\ref{eq:c1}) to obtain cancellation of the $1/\e^2$ poles in the 
 $[(N^2-1)/(2N)]^2=C_F^2$ and $(N^2-1)/(2N)N_f=C_FN_f$  
colour structures. It can be 
seen that the $q\bar q gg$ final state also yields a term canceling 
the contribution from the one-loop soft gluon current in the 
three-parton channel. Any remaining $1/\e$ poles are collected in 
\begin{equation}
\label{eq:htwor}
\bra{\cm^{(0)}}{\bom H}_R^{(2)}(\epsilon)\ket{\cm^{(0)}} 
=\frac{e^{\epsilon \gamma}}{4\,\epsilon\,\Gamma(1-\epsilon)} H_R^{(2)} 
\braket{\cm^{(0)}}{\cm^{(0)}} \;,  
\end{equation}
with
\begin{equation}
H^{(2)}_R = 2 H^{(2)}_{R,q}\;.
\end{equation}
The expression for $H^{(2)}_{R,q}$ in the \MSbar\ scheme reads, 
\begin{eqnarray}
H^{(2)}_{R,q}&=&
\left(\frac{51}{4}\zeta_3-{\frac {14315}{864}}- {\frac {49\pi^2}{288}}
\right)N^2
+\left(-\frac{105}{4}\zeta_3-{14731\over 432}+{13\pi^2\over 288}\right)
+\left(\frac{27}{2}\zeta_3-{561\over 32} + \frac{\pi^2}{8}\right){1\over
N^2}\nonumber \\
&&
+\left({11\pi^2\over 144}-{35\over 108}\right){(N^2-1)N_F\over N}\;.
\end{eqnarray}
The single pole constants from single virtual and double real radiation 
add up to the single pole constant from the double virtual contribution,
\begin{equation}
 H^{(2)}_{V,q} +  H^{(2)}_{R,q} =  H^{(2)}_{q}.
\end{equation}

Subtraction of the pole terms leaves
\begin{eqnarray}
\frac{\Finite^{(0 \times 0)}_{q\bar q (ij)}}{4Nq^2} 
&=& \left(N-\frac{1}{N}\right)
\Bigg[     N \left(
            \frac{1264873}{10368}
          + \frac{19\pi^2}{108}
          - \frac{4217}{72}\,\zeta_3
          - \frac{437\pi^4}{1440}
          \right)\nonumber \\
&&       +  \frac{1}{N}\left(
          - \frac{10637}{128}
          + \frac{2\pi^2}{3}
          + \frac{135}{4} \,\zeta_3
          + \frac{7\pi^4}{18}
          \right)
       +  N_f \left(
          - \frac{7883}{1296}
          - \frac{41\pi^2}{216}
          + \frac{133}{36}\,\zeta_3
          \right)\Bigg]. 
\end{eqnarray}

\section{Structure of infrared cancellations}
\setcounter{equation}{0}
\label{sec:ircancel}

In any infrared safe physical observable, such as the two-jet cross section 
or the inclusive hadronic production rate, one observes the cancellation of 
all infrared singularities present in individual partonic channels after 
adding all channels contributing to the same (infrared safe)
final state~\cite{kln}. At NLO, this cancellation is obvious in 
\begin{equation}
\Poles_{q\bar q}^{(1 \times 0)}+
 \Poles^{(0 \times 0)}_{q\bar q g}=0\;.
\end{equation}
At NNLO, the cancellation reads
\begin{equation}
\Poles_{q\bar q}^{(2 \times 0)}+
\Poles_{q\bar q}^{(1 \times 1)}+ \Poles^{(1 \times 0)}_{q\bar q g}
+\Poles^{(0 \times 0)}_{q\bar q (ij)} =0\;.
\end{equation}
In the computation of any jet observable, only the finite terms from 
each partonic channel will appear. 

As a check, we compute the perturbative 
corrections to the hadronic $R$-ratio, which is obtained by simply setting 
all measurement functions ${\cal F}_{J}^{(m)}$ to unity,
\begin{eqnarray}
\hat{R} &=& \frac{R_{{\rm had}}}{R^{{\rm tree}}_{{\rm had}}} \\
&=&
1 + \left(\frac{\alpha_s}{2\pi}\right) \, 
\left(\frac{{\cal T}^{(4)}_{q\bar q} (q^2) + 
{\cal T}^{(4)}_{q\bar qg} (q^2) }{{\cal T}^{(2)}_{q\bar q} (q^2) }
\right) \nonumber \\
& & + \left(\frac{\alpha_s}{2\pi}\right)^2 \, 
\left(\frac{{\cal T}^{(6)}_{q\bar q} (q^2) + 
{\cal T}^{(6)}_{q\bar qg} (q^2) + 
{\cal T}^{(6)}_{q\bar qq\bar q} (q^2) + 
{\cal T}^{(6)}_{q\bar qq'\bar q'} (q^2)+
{\cal T}^{(6)}_{q\bar qgg} (q^2)}{{\cal T}^{(2)}_{q\bar q} (q^2) }
\right)  \\
&=&
1 + \left(\frac{\alpha_s}{2\pi}\right) \, 
\left(\frac{
\Finite_{q\bar q}^{(1 \times 0)}
+ \Finite^{(0 \times 0)}_{q\bar q g}}{4Nq^2}
\right) \nonumber \\
& & + \left(\frac{\alpha_s}{2\pi}\right)^2 \, 
\left(\frac{
\Finite_{q\bar q}^{(2 \times 0)}+
\Finite_{q\bar q}^{(1 \times 1)}+ \Finite^{(1 \times 0)}_{q\bar q g}
+\Finite^{(0 \times 0)}_{q\bar q (ij)}}{4Nq^2}
\right)  \\
&=& 1 + \left(\frac{\alpha_s}{2\pi}\right) \, \left(\frac{N^2-1}{2N} \right)\, \frac{3}{2}
\nonumber \\ && 
+ \left(\frac{\alpha_s}{2\pi}\right)^2 \, \left(\frac{N^2-1}{2N}\right) 
\left[ N\left( \frac{243}{16} - 11 \zeta_3 \right)
+ \frac{1}{N} \, \frac{3}{16} 
+N_f\left(-\frac{11}{4}+2\zeta_3\right) \right]\;. 
\end{eqnarray}
This result is in agreement with the 
literature~\cite{chetyrkin}, thus providing a further
strong check on the correctness of all terms computed here.

\section{Summary and Conclusions}
\setcounter{equation}{0}
\label{sec:conc}

In this paper, we analytically 
examined the infrared singularity structure of 
two-jet production in $e^+e^-$ annihilation. 

For this purpose, we developed a subtraction formalism including
double real radiation at tree level and single real radiation at one loop.
In the case of the two-jet production process considered here, the 
subtraction terms coincide with the full tree level four-parton and 
one-loop three-parton matrix elements. The phase space structure of the 
subtraction extends the NLO dipole subtraction formalism~\cite{cs} to 
NNLO. In particular, the subtraction terms are integrated (both numerically 
and analytically) over the inclusive phase space. For single-particle 
subtraction at tree level and one loop, this procedure is 
already sufficient to transfer all infrared singularities from the numerical
integration (involving the jet definition) to an analytic expression which is
canceled against infrared poles from virtual corrections. This procedure 
has however to be extended in the case of double-particle subtraction. Indeed, 
the inclusive integral of the subtraction term will yield infrared 
singularities from double real emission (which subtract the corresponding 
terms in the full tree level double real emission cross section) as well as
infrared singularities from single real emission. The latter have to be 
subtracted {\it from the subtraction term} in order to avoid spurious 
infrared poles in the cross section. The infrared singularities 
from this second level subtraction can in the present calculation 
be identified to correspond to three-jet final states, which are 
part of the inclusive four-parton cross section. They are canceled by explicit 
infrared poles present in the virtual one-loop single emission subtraction 
term, which is  integrated over the inclusive three-parton phase space. In
fact, this one-loop single emission subtraction term
compensates the infrared singularities
 of the one-loop three-parton contribution to two-jet final 
states, but also yields purely virtual infrared poles in the 
three-jet region. 

The infrared pole structure of each partonic final state contribution to 
two-jet final states is computed explicitly by analytical integration of the 
appropriate subtraction terms. To analyze the resulting infrared poles
and their cancellation among different contributions, we start from 
Catani's  infrared factorization formula for two-loop amplitudes~\cite{catani}
and identify individual terms in this formula with corresponding terms in 
the integrals of the tree-level double real emission and one-loop  
single real emission subtraction terms. It turns out that an
identification up to the double pole terms is possible, provided a
particular finite constant (which is irrelevant for the infrared structure of
the two-loop virtual corrections)
 is chosen for the infrared singularity operator. Besides these terms 
canceling with infrared singularities in the two-loop virtual corrections, we
identify  one particular contribution which cancels between double real
emission and one-loop single real emission. This contribution is found
to be proportional to the one-loop correction to the soft gluon 
current~\cite{cg}. These identifications may help to find a more detailed 
insight into the structure of NNLO
infrared singularities of processes with 
more than two jets in the final state. Once this is accomplished, it appears 
feasible to construct NNLO infrared subtraction terms similar to 
the ones proposed here for processes 
with higher partonic multiplicity and/or combined emission in initial and 
final state.

\section*{Acknowledgments}
The authors 
would like to acknowledge the
hospitality of the Workshop 
``Physics at TeV colliders'' (Les Houches, France,
May 2001), where this work was initiated.

We would like to thank Gudrun Heinrich for pointing our attention to several
misprints in an earlier version of the manuscript.

This research was supported in part by the Swiss National Funds 
(SNF) under contract 200021-101874, by the National Science 
Foundation under Grant No.\ PHY99-07949, 
 by the UK Particle Physics and Astronomy  Research Council and by
the EU Fifth Framework Programme  `Improving Human Potential', Research
Training Network `Particle Physics Phenomenology  at High Energy Colliders',
contract HPRN-CT-2000-00149.

\begin{appendix}
\renewcommand{\theequation}{\mbox{\Alph{section}.\arabic{equation}}}
\section{Master Integrals}
\setcounter{equation}{0}
\label{app:ps}
All loop and phase space integrals in Section~\ref{sec:nnlo} were carried 
out by first reducing them to a small set of master integrals. This 
reduction is based on an iterative solution of integration-by-parts 
identities~\cite{chet} and is explained in detail in~\cite{laporta,gr,ggh}. 
For reference, we collected the analytic expressions (which can be found 
in various places in the literature~\cite{kl,3jmaster,ggh}) for all 
 master integrals in this appendix.

\subsection{Virtual two-loop corrections}
The virtual two-loop vertex master integrals were first derived 
in~\cite{kl} in the context of the calculation of the two-loop quark form 
factor~\cite{vanneerven}. Factoring out a common
\begin{equation}
S_\Gamma = \left(\frac{(4\pi)^{\e}}{16\pi^2\,\Gamma(1-\e)
}\right)^2\, ,
\end{equation}
they read
\begin{eqnarray}
A^2_{2,{\rm LO}} &=& 
\int \frac{\d^d k}{(2\pi)^d} \,\int \frac{\d^d l}{(2\pi)^d} \, 
\frac{1}{k^2(k-p_1-p_2)^2 l^2(l-p_1-p_2)^2}
\nonumber \\ &=&S_\Gamma \,\left(-q^2\right)^{-2\e}\,
 \frac{ \Gamma^2(1+\e)\Gamma^6(1-\e) }{
 \Gamma^2(2-2\e)}\frac{-1}{\e^2} \;,
\\
A_3 &=&\int \frac{\d^d k}{(2\pi)^d} \,\int \frac{\d^d l}{(2\pi)^d} \, 
\frac{1}{k^2 l^2 (k-l-p_1-p_2)^2}
\nonumber \\ 
&=& S_\Gamma\,  \left( -q^2 \right)^{1-2\e}\, 
\frac{\Gamma(1+2\e)\Gamma^5(1-\e)}{
\Gamma(3-3\e)} 
\frac{-1}{2(1-2\e)\e}\;,
\\
A_4 &=&\int \frac{\d^d k}{(2\pi)^d} \,\int \frac{\d^d l}{(2\pi)^d} \, 
\frac{1}{k^2 l^2 (k-p_1-p_2)^2 (k-l-p_1)^2}
\nonumber \\ 
&=& S_\Gamma \,\left(-q^2\right)^{-2\e}\,
 \frac{ \Gamma(1-2\e)\Gamma(1+\e)\Gamma^4(1-\e) 
 \Gamma(1+2\e)}{
\Gamma(2-3\e)}\frac{-1}{2(1-2\e)\e^2} \;,
\\
A_6 &=&\int \frac{\d^d k}{(2\pi)^d} \,\int \frac{\d^d l}{(2\pi)^d} \, 
\frac{1}{k^2 l^2 (k-p_1-p_2)^2 (k-l)^2 (k-l-p_2)^2 (l-p_1)^2}
\nonumber \\ 
&=&  S_\Gamma \,\left(-q^2\right)^{-2-2\e}\, \left[
 -\frac{1}{\e^4} + 
\frac{5\pi^2}{6\e^2} + \frac{27}{\e}\zeta_3 + \frac{23\pi^4}{36} 
+ {\cal O}(\e) \right]\;.
\end{eqnarray}

\subsection{Three-particle phase space integrals of one-loop corrections}
The inclusive three- and four-particle phase space integrals 
appear in this calculation in the form of dipole and tripole 
phase space integrals. We therefore factor out not only the 
usual normalization factor of dimensional regularization, but also 
the volume of the two-particle phase space into
\begin{equation}
S_{\Gamma,2} = P_2\, \left(\frac{(4\pi)^{\e}}{16\pi^2\,\Gamma(1-\e)
}\right)^2\, .
\end{equation}
The master integrals appearing as 
three-particle phase space integrals of one-loop matrix elements
are
\begin{eqnarray}
V_{5,a} &=& \Re \left[ -i \int \d \Phi_3 
\; \int \frac{\d^d k}{(2\pi)^d} \, \frac{1}{k^2 
(k-p_1-p_2-p_3)^2}\right] \nonumber \\
&=&S_{\Gamma,2}\,(q^2)^{1-2\e} \frac{\Gamma^6(1-\e)\Gamma(1+\e)}
{\Gamma(2-2\e)\Gamma(3-3\e)} \frac{-1}{\e} \Re (-1)^{-\e} \\
V_{5,b} &=&\Re \left[ -i \int \d \Phi_3 
\; \int \frac{\d^d k}{(2\pi)^d} \, \frac{1}{k^2 (k-p_1-p_3)^2 
}\right] \nonumber \\
&=& S_{\Gamma,2}\,(q^2)^{1-2\e} \frac{\Gamma^5(1-\e)\Gamma(1-2\e)\Gamma(1+\e)}
{\Gamma(2-2\e)\Gamma(3-4\e)} \frac{-1}{\e} \Re (-1)^{-\e}
\\
V_8 &=& \Re \left[ -i \int \d \Phi_3 \, \frac{1}{2 p_1 \cdot p_2} 
\; \int \frac{\d^d k}{(2\pi)^d} \, \frac{1}{k^2 (k-p_1)^2 (k-p_1-p_3)^2 
(k-p_1-p_2-p_3)^2}\right] \nonumber \\
&=& S_{\Gamma,2}\,(q^2)^{-2-2\e}\,\Bigg[ -\frac{5}{2\e^4}+\frac{9\pi^2}{2\e^2}
+\frac{89\zeta_3}{\e}+\frac{13\pi^4}{180}+ {\cal O}(\e)
\Bigg]\;.
\end{eqnarray}
The master integral $V_8$ was first derived in~\cite{ggh}.

\subsection{Four-particle phase space integrals}
Inclusive four-particle phase space integrals of tree level matrix elements 
can be expressed as linear combination of four master integrals, which were 
derived in~\cite{ggh}:
\begin{eqnarray}
R_4 & = & \int \d \Phi_4 = P_4 \nonumber \\
&=&
S_{\Gamma,2}\; (q^2)^{2-2\e}\, 
\frac{\Gamma^5(1-\e)\Gamma(2-2\e)}{\Gamma(3-3\e)
\Gamma(4-4\e)} \;, 
\\
R_6 & = & \int \d \Phi_4 \frac{1}{s_{134}s_{234}}
\nonumber \\
&=&
S_{\Gamma,2} \;
(q^2)^{-2\e}\,
\Bigg[ -1 + \frac{\pi^2}{6} + \e  \left(
-12+\frac{5\pi^2}{6}+9 \zeta_3 \right)\nonumber \\
   &&  \hspace{2cm} + \e^2  \left(-91+\frac{9\pi^2}{2}+45\zeta_3+\frac{61\pi^4}{180}\right)
+ {\cal O}(\e^3) \Bigg]\;,
\\
R_{8,a} & = & \int \d \Phi_4 \frac{1}{s_{13}s_{23}s_{14}s_{24}}\nonumber \\
&=&
S_{\Gamma,2} \;
(q^2)^{-2-2\e}\,
\Bigg[\frac{5}{\e^4}-\frac{20\pi^2}{3\e^2}
-\frac{126\zeta_3}{\e}+\frac{7\pi^4}{18} + {\cal O}(\e) \Bigg]
\;,
\\
R_{8,b} & = & \int \d \Phi_4 \frac{1}{s_{13}s_{134}s_{23}s_{234}}\nonumber \\
& = & 
S_{\Gamma,2} \;
(q^2)^{-2-2\e}\,
\Bigg[ \frac{3}{4\e^4}-\frac{17\pi^2}{12\e^2}
-\frac{44\zeta_3}{\e}-\frac{61\pi^4}{60}+ {\cal O}(\e)\Bigg]
\;.
\end{eqnarray}
\end{appendix}

\end{document}